\def\aa{{ A\&A}}
\def\aas{{ A\&AS}}

\def\aj{{AJ}}

\def\al{$\alpha$}
\def\bet{$\beta$}

\def\apj{{ApJ}}

\def\apjs{{ApJS}}

\def\asec{$^{\prime\prime}$}

\def\baas{{BAAS}}

\def\deg{$^{\circ}$}

\def\e#1{$\times$10$^{#1}$}
\def\etal{{et al. }}

\def\flux{ergs s$^{-1}$ cm$^{-2}$}

\def\kms{km\thinspaces$^{-1}$}
\def\lamb{$\lambda$}
\def\lum{ergs s$^{-1}$}

\def\mnras{{MNRAS}}

\def\nat{{Nature}}

\def\pasp{{PASP}}

\def\percm2{cm$^{-2}$}

\def\solum{L$_\odot$}

\def\gax    {${_>\atop^{\sim}}$ }

\def\kms    {~km~s$^{-1}$}
\def\refindent{\par\noindent\parskip=2pt\hangindent=3pc\hangafter=1 }

\def\>Z{$>$}
\def\<{$<$}

\def\simlt{\lower.5ex\hbox{$\; \buildrel < \over \sim \;$}}
\def\simgt{\lower.5ex\hbox{$\; \buildrel > \over \sim \;$}}
\def\sqr#1#2{{\vcenter{\hrule height.#2pt
      \hbox{\vrule width.#2pt height#1pt \kern#1pt
         \vrule width.#2pt}
      \hrule height.#2pt}}}
\def\mgii{\ion{Mg}{2}}

\def\oii{[\ion{O}{2}]}
\def\4363{[\ion{O}{3}]}
\def\heii{\ion{He}{2}}
\def\oiii{[\ion{O}{3}]}

\def\oi{[\ion{O}{1}]}

\def\nii{[\ion{N}{2}]}
\def\hei{\ion{He}{1}}
\def\sii{[\ion{S}{2}]}

\def\hii{\ion{H}{2}}

\def\farcs{\hbox{$.\!\!^{\prime\prime}$}}

\documentstyle[11pt,aaspp4]{article}
\received{}
\accepted{}

\slugcomment{To appear in {\it The Astrophysical Journal Supplement
Series}}

\begin{document}

\title{A Search for ``Dwarf'' Seyfert Nuclei. IV. Nuclei with Broad H\al\ 
Emission}

\author{Luis C. Ho}
\affil{Department of Astronomy, University of California, Berkeley, CA 
94720-3411}

\and

\affil{Harvard-Smithsonian Center for Astrophysics, 60 Garden St., Cambridge,
MA 02138\footnote{Present address.}}

\author{Alexei V. Filippenko}
\affil{Department of Astronomy, University of California, Berkeley, CA 
94720-3411}

\author{Wallace L. W. Sargent}
\affil{Palomar Observatory, 105-24 Caltech, Pasadena, CA 91125}

\and
\author{Chien Y. Peng}
\affil{Department of Astronomy, University of California, Berkeley, CA 
94720-3411}

\begin{abstract}

We present the results of an optical spectroscopic survey designed to search 
for low-luminosity, ``dwarf'' Seyfert nuclei in a magnitude-limited 
($B_T\,\leq$ 12.5 mag) sample of 486 bright, northern galaxies.  
Moderate-resolution spectra of exceptionally high quality were obtained in part
to detect broad H\al\ emission, similar in character to, but much weaker than, 
the broad permitted lines that define type 1 Seyfert nuclei.  One of the 
primary goals of the survey is to better quantify the faint end of the 
luminosity function of active galactic nuclei.  

This paper describes the subset of nuclei showing definite or probable 
evidence of broad H\al\ emission.  We outline the procedures for determining 
the presence of this elusive spectral feature, steps for its quantitative 
measurement, and the associated systematic errors.   Of the 211 emission-line 
galaxies classified as having Seyfert or LINER (low-ionization nuclear 
emission-line region) nuclei in our survey, the broad H\al\ line was detected 
with confidence in 34 objects, and with less certainty in another 12.  Most 
of the detections are reported for the first time, and the detection rate 
represents a lower limit to the true incidence of active nuclei harboring a 
broad emission-line region.  These statistics imply that broad-lined 
active nuclei are much more common than previously believed: they exist in 
at least 20\% of all galaxies spectroscopically classified as ``active'' 
and in more than 10\% of all luminous galaxies at the current epoch.

Excluding nine well-known Seyfert 1 nuclei, the broad H\al\ line of 
the remaining ``dwarf'' nuclei has a median luminosity of $\sim$1\e{39} 
\lum\ and a median full width at half maximum of $\sim$2200 \kms.  It 
constitutes approximately 30\% of the total emission of the H\al\ and 
\nii\ \lamb\lamb 6548, 6583 blend.  Several objects have broad H\al\ 
luminosities as low as (1--3)\e{38} \lum.

\end{abstract}


\section{Introduction}

Active galactic nuclei (AGNs) manifest themselves over a wide range of 
luminosities.  Although the nuclei of Seyfert 1 galaxies have $L_B\,\approx\,
10^{8} - 10^{11}$ \solum\ ($M_B\,\approx$ --16 to --23 mag; Weedman 1976), 
it is generally accepted that they represent the phenomenon associated 
with rare (at the present time) and more luminous QSOs, whose power output is
 $L_B\,\approx\, 10^{11} - 10^{14}$ \solum\ ($M_B$ = --23 to --30 mag;
Schmidt \& Green 1983).  The basis for this belief rests on two main 
observational facts.  On the one hand, Seyfert 1 nuclei and QSOs have 
very similar spectroscopic properties (Weedman 1976); on the other, it 
has become increasingly apparent that many low-redshift QSOs reside in 
structures photometrically resembling galaxies (Wyckoff \etal 1980; Malkan, 
Margon, \& Chanan 1984; McLeod \& Rieke 1994a, b; Boyce \etal 1996; Bahcall 
\etal 1997), and spectroscopic confirmation that the extended nebulosities 
around QSOs come from starlight has been possible in some instances (Wyckoff 
\etal 1980; Boroson, Oke, \& Green 1982; Balick \& Heckman 1983; Miller, 
Tran, \& Sheinis 1996).  

The degree to which the AGN phenomenon extends to luminosities fainter than 
those of ``classical'' Seyferts has not yet been established.  Several optical 
spectroscopic surveys of nearby galaxies [see Filippenko (1989) 
and Ho (1996) for reviews] corroborate that emission-line nuclei exhibiting 
excitation patterns indicative of nonstellar photoionization are quite common 
in the present epoch.  The extensive survey recently completed by Ho, 
Filippenko, \& Sargent (1995), for example, revealed an astonishingly large  
population of galaxies with ``active'' nuclei: nearly half of all galaxies 
brighter than $B_T$ = 12.5 mag can be classified as low-ionization nuclear 
emission-line regions (LINERs; Heckman 1980) or Seyferts (Ho 1995, 1996; Ho, 
Filippenko, \& Sargent 1997b, hereafter Paper V).  The nonstellar continuum 
and emission-line luminosities of these nuclei, however, can be orders of 
magnitude fainter than those of conventionally studied AGNs.  The 
LINER/Seyfert nucleus of NGC 3031 (M81), for instance, has $M_B$ = --11.6 mag 
and a broad H\al\ luminosity of only 1.8 $\times\,10^{39}$ \lum\ (Ho, 
Filippenko, \& Sargent 1996), while in the extreme case of NGC 4395, its 
Seyfert 1 nucleus has $M_B$ = --9.8 mag and $L$(H\al) = 1.2 $\times\,10^{38}$ 
\lum\ (Filippenko \& Sargent 1989).  

The exceptionally low luminosities of these sources have presented a serious 
observational challenge and have proven to be a major obstacle to furthering 
our understanding of them.  Whether all the mildly ``active'' nuclei are 
low-power but genuine AGNs, and, if so, how these fit into the standard AGN 
paradigm remain outstanding unanswered questions.  Particularly troublesome are 
LINERs, which make up the bulk of the candidate AGNs, because their 
association with a nonstellar energy source is but one of several possible 
interpretations (Ho, Filippenko, \& Sargent 1993; Filippenko 1996).  The 
fundamental difficulty in determining the physical nature of low-luminosity 
AGNs stems from the extremely low contrast between the nucleus and the 
surrounding bulge.  At optical wavelengths, the conventional telltale signs of 
AGN-related activity such as the presence of a starlike nucleus, a blue, 
featureless continuum, or rapid variability, generally cannot be seen against 
the glare of the host galaxy.  One possible exception may be the detection of 
broad emission lines.  QSOs and luminous Seyfert 1 nuclei distinguish 
themselves unambiguously by their characteristic broad [full width at half 
maximum (FWHM) of a few thousand \kms] permitted lines (Khachikian \& Weedman 
1974).  If such broad lines were to be found in low-luminosity AGNs, then one 
could argue for an underlying physical connection between the two luminosity 
regimes.  Since the strongest permitted line at optical wavelengths is 
expected to be H\al, searching for broad H\al\ emission is a promising method 
to constrain the origin of the widespread nuclear activity found in 
spectroscopic surveys of nearby galaxies.

Weak, broad H\al\ emission has long been known to be present in some nearby 
AGNs.  Peimbert \& Torres-Peimbert (1981), for example, noticed that the 
nucleus of NGC 3031 emits broad H\al, and several other isolated cases have 
been noticed from time to time (e.g., Stauffer 1982; Keel 1983).  Because of 
the formidable challenge in detecting this weak spectral feature, what has 
been lacking so far is a {\it systematic} study that can provide meaningful 
statistics on the incidence of broad H\al\ emission.  We have recently 
completed a survey aimed at achieving this goal.  Using the Palomar 5-m Hale 
telescope, long-slit optical spectra were obtained of the nuclei of nearly 
500 northern galaxies (Filippenko \& Sargent 1985, hereafter Paper I).  The 
sample is defined to be all galaxies listed in the Revised Shapley-Ames 
Catalog of Bright Galaxies (Sandage \& Tammann 1981) with $\delta\,>$ 0\deg\ 
and $B_T\,\leq$ 12.5 mag. Paper II (Ho \etal 1995) presents the spectral 
atlas of the survey and discusses the observational parameters and data 
reduction; Paper III (Ho, Filippenko, \& Sargent 1997a) gives the detailed line 
measurements and object classifications; and Paper V discusses the overall 
detection rates of the different classes of emission-line nuclei, their 
dependence on the host galaxy morphology and luminosity, and issues 
concerning completeness.  

Here we summarize the overall properties of the subsample of emission-line 
nuclei showing broad H\al\ emission.  We describe the procedures necessary for 
confident determination of the presence of this weak, elusive spectral 
feature, steps for its quantitative measurement, and the associated systematic 
errors (\S\ 2).  Plots of the data are provided for all candidates with 
positive detections, and (to illustrate our methods) for several others whose 
broad H\al\ line is either more ambiguous or absent (\S\ 3).  Section 4 
discusses some of the overall implications of this work.

\section{Line Deblending Techniques}

\subsection{General Constraints}

A number of systematic effects plague the detection of faint, broad H\al\ 
emission.  The most serious of these is contamination from late-type stars 
which dominate the integrated optical spectra of the centers of early-type 
spirals, the preferred hosts of active nuclei (e.g., Keel 1983; Paper V).  As 
illustrated in Paper I, the continuum and spectral features of K and M giants 
can easily mimic a broad emission hump near the wavelength of H\al.  Thus, 
detection of broad H\al, especially if weak, requires careful removal of the 
underlying starlight.  Although this affects less severely sources with strong 
broad H\al\ emission, quantitative measurements still require its 
consideration.  The details of the starlight subtraction for all the objects 
analyzed in this paper are given in Paper III.

For the line fitting procedure itself, it is necessary to impose some 
constraints in order to reduce the number of free parameters, as the H\al\ + 
\nii\ region (interchangeably called the ``complex'' in subsequent 
discussions) is generally quite complicated.  In addition to a possible broad 
H\al\ component, there are narrow lines of H\al\ and \nii\ \lamb\lamb 6548, 
6583, which rarely show Gaussian profiles.  The narrow lines generally have 
faint, extended wings near their bases, some degree of asymmetry, velocity 
blending among the three lines, and, occasionally, multiple velocity peaks.  
Any of these effects, alone or combined, can mimic or hide a weak broad H\al\ 
component.  Consequently, we have taken great care to examine all known 
systematic effects.  Three methods, described below, were explored; each has 
its own merits and weaknesses depending on the object in question, and, in some 
cases, the final line parameters quoted were obtained by averaging the results 
of all three.

We assume that the H\al\ and \nii\ lines, and likewise the \sii\
\lamb\lamb 6716, 6731 lines, are separated by their laboratory wavelengths in 
the rest-frame spectra.  The intensity ratio of \nii\ \lamb 6583 to \lamb 
6548 is fixed at the theoretical value of 2.96, and we require that the two 
\nii\ line profiles be identical.  Since the \sii\ doublet suffers from 
somewhat less confusion, in most cases a model for the narrow lines is derived 
from \sii.  (A different procedure is adopted for spectra whose
signal-to-noise ratio [S/N] in the \sii\ region is poor; see \S\ 3.)
We then use the model to deblend the more complicated H\al\ + \nii\
region, usually assuming that all the narrow lines have the same profile, but
occasionally having to relax this constraint, especially for H\al.  Although
it is well known that the forbidden lines in AGNs can have different profiles
depending on their respective critical densities (Filippenko \& Halpern 1984; 
Filippenko 1985; De Robertis \& Osterbrock 1986), \nii\ often has a line 
profile that closely matches that of \sii, despite having a significantly 
larger critical density (Filippenko \& Halpern 1984; Filippenko 1985; 
Filippenko \& Sargent 1988).  Perhaps the \sii\ and \nii\ lines are dominated 
by emission from a relatively low-density reservoir of gas outside the 
stratified part of the narrow-line region.

\subsection{Automated Fitting}

The \sii\ doublet usually can be fitted by a combination of several Gaussians, 
thereby yielding a set of analytic components which can then be applied to 
the narrow lines of H\al\ and \nii.  The number of Gaussians for each of the 
\sii\ lines, $N$, varies from 1 to as many as 4 or more, depending on the 
complexity of the profile.  (Note that we do not necessarily ascribe physical 
significance to each individual Gaussian; this is merely a convenient fitting 
procedure.)  Whenever possible, we make the first-order assumption that 
the components have the same intensity ratios for the two \sii\ lines; this 
constraint must be relaxed in cases where the electron densities of the 
components are not similar.  Figure 1 illustrates these two scenarios.

Once a \sii\ model is derived, we fit the H\al\ + \nii\ complex using $3N+1$ 
Gaussians, $N$ components for each of the narrow lines and one for broad 
H\al.  As explained above, we force the \nii\ lines to have identical 
profiles and a fixed intensity ratio, while the narrow component of H\al\ is
allowed to vary in its profile (but not in central wavelength) if necessary. 
Figure 2{\it a} shows an example of this procedure for the case of NGC 4278.

\subsection{Analytic and Synthetic Profile Scaling}

The broad H\al\ profiles of some objects, notably most of the classical 
Seyfert 1 nuclei included in our sample, have highly irregular shapes which 
cannot be fitted with a single Gaussian.  Imposing such a constraint would
lead to clearly erroneous fits for the narrow lines, while increasing the
number of Gaussian components to better match the broad line would allow too
much freedom for the final solution of the narrow lines (at least when the 
broad and narrow components are not well separated).  Under these
circumstances, it is best simply to subtract appropriately scaled versions of
the \sii\ profile in order to remove the contribution of the narrow lines.
We require that the \nii\ lines have the correct intensity ratio and that the
final broad H\al\ profile be as smooth and continuous as possible.  While 
admittedly somewhat subjective, in some instances such a procedure offers the 
only practical solution.  

The model \sii\ profile can be 
obtained in two ways.  The first method, hereafter referred to as ``analytic 
profile scaling,'' is identical to that of the initial phase of ``automated 
fitting'' --- namely, the best-fitting model for \sii\ is obtained by finding a
suitable combination of Gaussians (Fig. 2{\it b}).  A second method, which
we will call ``synthetic profile scaling,'' constructs an empirical \sii\ 
profile from the observed blend.  As discussed by Filippenko \& Sargent 
(1988), for example, a symmetric \sii\ profile can be created by reflecting 
either the unblended blue half of the \lamb 6716 line or the red half of 
\lamb 6731.  Likewise, for obviously asymmetric lines, the blue half of 
\lamb 6716 can be combined with the suitably scaled red half of \lamb 6731 
(Fig. 2{\it c}).  Synthetic profile scaling can be easily implemented if the 
data have high S/N and if the \sii\ lines are reasonably resolved; if the two 
\sii\ lines severely overlap with each other, an iterative procedure can be 
attempted (e.g., Shields \& Filippenko 1990), although the uniqueness of the 
resulting model profile then becomes difficult to ascertain.

\subsection{Uncertainties and Systematic Errors}

We performed simulations to gauge the reliability of our automated fitting 
procedure; of the three procedures described above, this one proved to 
be most useful for most of the survey.  The tests were done using a high S/N 
spectrum known in advance to have narrow lines with very similar profiles and 
to lack any visible broad H\al\ emission.  We chose the spectrum of NGC 7217.  
Next, a grid of synthetic spectra was generated by introducing broad Gaussian 
components of various widths and amplitudes centered at the wavelength of 
H\al.  To better represent the average quality of the data, we added Poisson 
noise to the spectra, and we introduced small fluctuations of random amplitude 
and frequency to the continuum level to mimic imperfections as might arise 
from the starlight subtraction process.  We parameterized the broad emission 
line with FWHM = 20, 40, 60, and 80 \AA\ and $f_{blend}$ = 10\%, 20\%, 40\%, 
60\%, and 80\%, where $f_{blend}$ is the fraction of the flux in the 
H\al\ + \nii\ complex contributed by the broad H\al\ line; these values of 
FWHM and $f_{blend}$ roughly bracket the range of values actually measured 
in the survey galaxies.  We then constructed a model from the \sii\ doublet 
and attempted to recover the broad H\al\ line.  

The results of these simulations (Fig. 3{\it a}) indicate that the input FWHM 
and $f_{blend}$ parameters of the broad H\al\ line can be recovered to 
within $\sim$10\% provided that FWHM \gax 10 \AA\ and $f_{blend}$ \gax 20\% 
for the case where the spectral resolution at H\al\ is assumed to be 2.5 \AA, 
the nominal value for our survey.
As to be expected, for a fixed line width, the errors become progressively 
larger when the broad line has low contrast with respect to the H\al\ + \nii\ 
blend ($f_{blend}$ small).  Conversely, for a fixed relative line strength, 
it is more difficult to recover a line with a larger breadth because it 
becomes increasingly washed out into the continuum.  So long as the 
\sii, H\al, and \nii\ lines have similar profiles, and provided that the 
S/N of the data, the quality of the starlight subtraction, and the spectral 
resolution are relatively high (i.e., comparable to those of our survey), we 
believe that our method can reliably extract a weak broad H\al\ component with 
FWHM $\approx$ 1000--3000 \kms\ even if it comprises only about 20\% of the 
flux of the H\al\ + \nii\ blend.  From the H\al\ line fluxes and \nii/H\al\ 
ratios reported in Paper III, the typical H\al\ + \nii\ blend for the AGNs in 
our sample has a flux of $\sim$3\e{-14} \flux; thus, we conservatively 
estimate that our flux detection limit for the broad H\al\ line is 
$\sim$6\e{-15} \flux.  This corresponds to an H\al\ luminosity of 3\e{38} 
\lum\ for the median distance of 20.6 Mpc for the AGN sample (Paper V).  In 
a few cases (e.g., see NGC 4388 in \S\ 3), even fainter levels can be 
reached.

For the benefit of workers who may only have access to data of lower spectral 
resolution, we repeated the simulations on spectra which were degraded to a 
resolution (at H\al) of 5 \AA\ (Fig. 3{\it b}) and 10 \AA\ (Fig. 3{\it b}).  
Not surprisingly, the errors are larger in these cases, but evidently it 
is still possible to detect broad H\al\ under most conditions, provided, of 
course, that the data have sufficiently high S/N and that starlight subtraction 
has been taken properly into account.

In some circumstances, discussed in detail where relevant in \S\ 3, our
techniques of profile fitting do not lead to unique solutions. 
Moreover, the broad H\al\ fluxes derived for a few objects can differ by large 
factors ($\sim$2) depending on the technique used, and it is not always clear 
which of these is the most reliable.  This problem most severely affects 
objects with a weak broad-line component.

One of the most important potential sources of systematic error stems from the 
assumptions made during the line-fitting process concerning the intrinsic 
profiles of the narrow components of H\al\ and \nii.  We always begin with 
the conservative and empirically supported (e.g., Filippenko \& Halpern 1984; 
Filippenko 1985; Filippenko \& Sargent 1988) hypothesis that the profiles of 
these lines can be approximated by that of at least one of the \sii\ lines.  
If all or most low-luminosity AGNs obey a strict line width versus 
critical density relation, \nii\ should have a broader profile than \sii.  
Although this appeared not to hold in the few objects analyzed in the 
above-mentioned studies, it should be noted that minor differences in the 
widths of the {\it wings} of the narrow lines would have escaped notice in 
the objects examined.  In the case of NGC 3031 (see \S\ 3 and Filippenko 
\& Sargent 1988), for instance, the broad component of H\al\ is sufficiently 
strong and non-Gaussian in shape that it would be difficult to perceive subtle 
mismatches among the profiles of narrow H\al, \nii, and \sii.  Indeed, as 
described in \S\ 3, profile differences among these three lines clearly exist 
in some objects.  An alternative strategy might be to use instead  \oiii\
\lamb 5007 as the model.  However, this does not alleviate the difficulty, 
since the critical density of \oiii\ is larger than that of \nii; moreover, 
although \oiii\ is usually quite strong, the blue spectra in our survey have 
lower spectral resolution (by about a factor of 2) and generally lower S/N 
than the red spectra.

Although the profiles of narrow H\al\ and \nii\ closely match each other in 
the majority of objects, exceptions do exist.  These can arise in one of two 
ways.  First, the height, and hence the width, of the narrow component of H\al\ 
emission critically depends on the accuracy with which the underlying H\al\ 
absorption line has been removed during the process of starlight subtraction.
Second, genuine profile differences {\it do} exist in objects with composite 
nuclei (e.g., see NGC 4303 in Paper I).  Mixing an \hii\ region component with 
an AGN component changes the line profiles of the two lines because the 
\nii/H\al\ ratio differs for the two excitation mechanisms.

Another source of uncertainty may enter through errors in the determination of 
the spectrograph response function, which directly affects the derived 
continuum shape.  In general, the response function for our data is known 
quite accurately, as judged by the calibrated continuum energy distributions 
of the standard stars observed along with the program galaxies.  
However, temporal variations in the spatial focus of the spectrograph, which 
can introduce low-frequency flux undulations in the spectra, occasionally 
plagued some of our data, particularly those taken with the red CCD camera.  
As discussed in Paper II, we took great effort to rectify this 
problem during the calibration process, and we believe that the broad H\al\ 
measurements in this paper should not be adversely affected.

Finally, systematic errors can enter during the starlight subtraction phase 
of the data analysis.  Mismatch of the overall continuum shape between the 
object spectrum and the absorption-line template, as might arise from
imperfect calibration, differences in reddening, or a true discrepancy in 
the stellar population, will result in artificial structure in the residual 
emission-line spectrum.  Our procedure for the starlight subtraction 
(Paper III), however, explicitly removes any gross differences in the general 
continuum shape between the object and the template.  An imperfect fit of the 
stellar population can also introduce, in one of two ways, a discrete feature 
mimicking broad H\al\ emission.  First, as already mentioned in \S\ 2.1, 
the continuum of late-type giant stars exhibits a broad maximum near the 
wavelength of H\al\ (see Fig. 12 in Paper I) that can easily be mistaken 
for a broad H\al\ emission line.  Our starlight subtraction should be 
sufficiently accurate that this effect is probably minor.  Second, if the 
underlying stellar population has a significant contribution from A-type 
stars, it is possible to generate an artificial ``bump'' resembling broad H\al\ 
emission.  Since the wings of the Balmer absorption lines in most subclasses 
of A stars are widened considerably by pressure broadening, oversubtraction 
of the A-star component in the template spectrum results in a broad peak 
centered at the expected position of each Balmer line.  Similarly, 
undersubtraction generates a negative depression (or ``bowl'') in the local 
continuum surrounding the Balmer emission line.  These signatures are easily 
observable at H\bet, where blending with neighboring emission lines is 
rarely an issue.  Since in our sample genuine broad H\bet\ emission is 
not observed except in the previously known Seyfert 1 nuclei and in a small 
number of the brightest other Seyfert 1s, the flatness of the local continuum 
near H\bet\ serves as an excellent gauge of how well the young stellar 
population has been removed (see also discussion in Paper III).

In the next section we try to fold these complications, which are difficult 
to quantify rigorously, into our evaluation of the presence or absence of 
broad H\al\ emission.

\section{Results}

Here we summarize the results of the profile fitting.  All objects with
definite or probable detections of broad H\al\ are discussed individually, 
with the accompanying fits shown to convey the maximum
amount of information.  To illustrate some technical issues, we also show
a number of examples of null detections.  Table 1 summarizes several parameters
of interest for the objects in which broad H\al\ has been definitely or 
probably detected.  Note that the parameters of the broad H\al\ line given in 
Table 1 refer only to observations taken at a single epoch.  Several objects 
are known to show variable broad-line emission, and these are highlighted 
below. In a few objects, mainly the well-known Seyfert 1 nuclei 
in the sample, a broad component is easily visible in permitted lines other
than H\al, and we point these out accordingly.

{\it NGC 266.} --- A broad H\al\ component is obvious even before profile 
decomposition, as evidenced by the extra ``bump'' between narrow
H\al\ and \nii\ \lamb 6583 (Fig. 4{\it a}).  Since the rather noisy \sii\ 
region could not be fitted to give a reliable model for the narrow lines, 
we resorted to fitting the H\al\ + \nii\ complex independently.  Each of the 
narrow lines required two components, as they are slightly asymmetric to the 
blue, but we forced all three lines to have identical profiles.  The resulting 
narrow-line profile, when applied to the \sii\ lines, provided a reasonable 
fit.  The broad H\al\ line, with FWHM $\approx$ 1350 \kms\ and $f_{blend}$ 
$\approx$ 30\%, appears to be redshifted by $\sim$350 \kms\ with respect to 
the narrow component of H\al.  The latter effect may be an artifact of the 
uncertain continuum level, as the negative dip shortward of \nii\ \lamb 6548 
probably influences the derived central wavelength of the broad H\al\ feature.

{\it NGC 315.} --- As noted previously by Ho \etal (1993), a broad H\al\ 
component is present in this radio galaxy with twin radio jets (Giovannini, 
Feretti, \& Comoretto 1989), but its strength turns out to be considerably 
weaker ($f_{blend}\,\approx$ 29\%) than what might be perceived by eye (Fig. 
4{\it b}).  The narrow lines have very broad, non-Gaussian wings, easily 
discernible, for instance, in the \sii\ lines.  Unfortunately, the model 
for the \sii\ lines is not well determined because \sii\ \lamb6731 is partly 
redshifted out of our spectral range.

{\it NGC 660.} --- This transition object (LINER+\hii\ nucleus) may contain 
extremely weak broad H\al\ emission, but its reality is suspect because of a 
possible template mismatch.  As the narrow lines have unusually small widths 
(FWHM $\approx$ 200 \kms), very faint emission between H\al\ and \nii\ 
is discernible.  When fitted with a single Gaussian, the broad component
contains about 8\% of the total flux of the blend (Fig. 4{\it c}).
 
{\it NGC 841.} --- The \sii\ lines are too noisy to provide a reliable
model; however, the narrow lines of H\al\ and \nii\ are fairly symmetric
and can be fitted with single Gaussians.  A four-component fit to the complex
(Fig. 4{\it d}) reveals a sizable broad H\al\ line ($f_{blend}\,\approx$
39\%).  The continuum surrounding the line complex unfortunately has large
residuals, making the fit somewhat uncertain.

%{\it NGC 877.} --- WR mismatch?

{\it NGC 1052.} --- This historically important object inspired much of the 
early discussion on the relative contribution of shocks and photoionization to
the excitation of LINERs (Fosbury \etal 1978; Keel \& Miller 1983; Rose \&
Tripicco 1984).  We confirm the presence of broad H\al\ emission ($f_{blend}\,
\approx$ 32\%; FWHM $\approx$ 1950 \kms) suspected in Paper I.  The evident 
asymmetry of the narrow lines, coupled with substantial rotational broadening 
(FWHM $\approx$ 480 \kms), conspire to make the fitting rather tricky.  
Nevertheless, a reasonably good fit can be achieved using the model derived 
from the \sii\ lines (Fig. 4{\it e}), and all three fitting methods give 
consistent results for the broad H\al\ flux.  The detection of broad H\al\ 
furnishes strong evidence that the nucleus of NGC 1052 should indeed be 
regarded as active.

{\it NGC 1068.} --- Malkan \& Filippenko (1983) first remarked that the 
Seyfert nucleus of NGC 1068 may have weak broad H\bet\ emission, although they 
did not explicitly model the line profile to measure the broad line.  The 
existence of a broad-line region (BLR) in this object was subsequently 
demonstrated conclusively by Antonucci \& Miller (1985), 
whose spectropolarimetric 
observations showed that the polarized spectrum closely resembled that of a 
typical Seyfert 1 nucleus.  Here, we wish to reexamine whether the broad 
component of H\bet\ and H\al\ can be detected in the total-light spectrum.
The kinematics and spatial distribution of the narrow-line region in NGC 1068 
are notoriously complicated.  It has long been known that the line-emitting 
material within the central several hundred parsecs consists of a number of 
distinct, high-velocity clumps (Walker 1968) which combine to produce very 
complex, exceptionally broad line profiles (e.g., Pelat \& Alloin 1980), 
that the integrated line profile changes dramatically over small angular 
scales (e.g., Alloin \etal 1983; Baldwin, Wilson, \& Whittle 1987), and that 
the line-intensity ratios of the different velocity components do not remain 
constant throughout the nuclear region (Baldwin \etal 1987; Cecil, Bland, \& 
Tully 1990).  Thus, we anticipate at the outset that many of the assumptions 
employed thus far in this study are unlikely to be applicable in NGC 1068.  

We begin with the H\bet\ line, since it is much less blended with surrounding 
lines than H\al.  Although the profiles of H\bet\ and \oiii\ differ in detail, 
previous studies indicate that these two lines roughly trace each other in 
overall shape (e.g., Cecil \etal 1990; Veilleux 1991), especially as seen in 
data with moderate spectral resolution.  We therefore constrain the 
narrow component of 
H\bet\ to have the same profile as that of \oiii.  An additional Gaussian was 
then introduced to test for the presence of possible broad H\bet\ emission.  
We leave the height of the Gaussian unconstrained but fix the FWHM to be 
3210 \kms\ and the line center redshifted by 600 \kms\ with respect to the 
narrow component; these parameters, taken from the spectropolarimetric study 
of Miller, Goodrich, \& Mathews (1991), pertain to the polarized broad H\bet\ 
profile from the nucleus, and should be identical to that of the hypothetical 
component viewed in total light.  We cannot allow too much freedom in the 
fit for the broad component because the extended bases of the narrow lines, 
especially the blue wing anticipated from \oiii\ \lamb 4959, can easily lead 
to an erroneous, even if formally acceptable, solution.  Our best fitting 
model for each of the \oiii\ lines (Fig. 5{\it a}) consists of a sum of five 
Gaussians: four have FWHM $\approx$ 400--600 \kms\ and contain about half of 
the flux, and another is considerably broader (FWHM $\approx$ 1900 \kms) and 
slightly blueshifted ($\sim$190 \kms) relative to the main narrow component.  
The broad, blueshifted feature identified here most likely corresponds to a 
similar component seen in many previous studies (e.g., Pelat \& Alloin 1980; 
Alloin \etal 1983; Caganoff \etal 1991) and lends confidence that our model 
for the narrow lines is qualitatively correct.  Under this assumption, the fit 
for the H\bet\ line {\it does} require a BLR component (see bottom panel of 
Fig. 5{\it a}).  Interestingly, the broad H\bet\ line has roughly the 
strength expected from the detailed model by Miller et al. for the scattered 
line flux of the nucleus.  The broad line we detect constitutes $\sim$19\% 
of the integrated flux of H\bet, and it has 1.8\% of the total intensity of 
\oiii\ \lamb 5007.  Although our observations of this source were not taken 
under photometric conditions, we can estimate the flux of the broad H\bet\ 
line as follows.  Assuming that the flux of \oiii\ \lamb 5007 is 
1.7\e{-11} \flux\ (Shields \& Oke 1975), as did Miller et al., the flux of 
broad H\bet\ should be 3.1\e{-13} \flux, in very close agreement with the 
value of 3.7\e{-13} \flux\ estimated by Miller et al.  

The H\al\ + \nii\ region is considerably more complicated.  In addition to 
the severe blending of the broad, complex lines, it appears that the 
intrinsic narrow-line profile of H\al\ does not match that of \nii.  If we 
construct an analytic model from \sii\ as usual (Fig. 5{\it b}), it is 
evident that the \sii\ lines have a much more extended blue tail than \oiii; 
a broad component is again evident (FWHM $\approx$ 2050 \kms), but it is 
significantly more blueshifted ($\sim$1000 \kms).  Thus, {\it a priori}, we 
expect that making the assumption that the profiles of \sii, \nii, and H\al\ 
are equal is invalid, since H\bet, which presumably should be nearly identical 
to H\al, is well fitted by \oiii.  On the other hand, we find that \oiii\ 
clearly does not match \nii, whereas \sii\ approximately does.  Accordingly, 
we constructed a model such that H\al\ and \oiii\ had identical profiles, 
and the same between \nii\ and \sii, and we once again added an extra Gaussian 
to test for the presence of broad H\al\ emission.  As in the case of H\bet, we 
fixed the position and the width of the Gaussian.  Under these assumptions, we 
{\it also} find evidence for a BLR component to H\al\ (Fig. 4{\it f}).  
Although the fairly large residuals indicate that our fit is far from perfect, 
omitting the BLR component results in much larger residuals.  As with the BLR 
component of H\bet, the broad H\al\ line contains $\sim$20\% of the total 
flux of H\al, and it has a sensible strength --- the ratio of broad 
H\al\ to broad H\bet\ is 7, close to values typically seen in Seyfert 1 
nuclei (e.g., Netzer 1990).  

To summarize, our profile analysis suggests that both the H\al\ and the H\bet\ 
emission lines display a broad component in the total-light spectrum.  We 
believe that this broad component arises from the BLR and is distinct from the
high-velocity emission associated with the narrow-line region.  Our 
measurement of this component, however, is by no means straightforward, and 
because of the simplifying assumptions that had to be adopted, we cannot be 
sure that our profile fits are unique.  Nevertheless, it is encouraging that 
the derived strength of the broad H\bet\ line agrees so well with the 
predictions of the Miller \etal (1991) model, and that the results of the 
H\al\ fit are consistent with those of H\bet.  It is of historical interest to 
note that, based on these results, NGC 1068 technically should {\it not} be 
classified as a Seyfert galaxy of type 2 but rather one of type 1.8 (weak broad 
H\al\ and H\bet; Osterbrock 1981).

{\it NGC 1161.} ---  Despite the poor S/N of the data, a broad component 
of H\al\ can be seen clearly in the starlight-subtracted spectrum even 
without profile fitting (Fig. 4{\it g}).  We obtained the final fit by 
repeated trials in which both the FWHM and central wavelength of the broad 
component were fixed at different input values, while its height and the
parameters of the narrow lines were varied.
 
{\it NGC 1167.} ---  The presence of faint wings near the base of the \nii\
lines led us to suspect in Paper I that broad H\al\ might be present in 
NGC 1167.  However, similar wings can be seen in the \sii\ profiles, and a 
careful fit of the H\al\ + \nii\ blend indicates that broad H\al\ is not 
present (Fig. 4{\it h}).
 
{\it NGC 1275.} --- As discussed in Paper I, the broad H\al\ component of 
NGC 1275 has extremely wide wings.  In addition to their unusually large 
breadth (FWHM $\approx$ 450 \kms\ for \nii\ and \sii), the narrow lines also 
have quite extended wings, making decomposition difficult.  Nevertheless, each 
\sii\ line can be represented by three Gaussians, the sum of which was then 
used to remove the narrow lines from the complex by analytic profile scaling 
(Fig. 14{\it a}); repeating the process using synthetic profile scaling gave 
virtually identical results.  The final broad H\al\ line has FWHM $\approx$ 
2750 \kms, full width near zero intensity (FWZI) $\sim$19,000 \kms, and 
contains 59\% of the flux of the entire blend.  A broad component has also 
been detected in H\bet, H$\gamma$, and \heii\ \lamb 4686.

{\it NGC 1358.} --- This object provides an excellent illustration of the 
potential ambiguities and limitations of our method.  If we make the usual 
assumption that the \sii\ lines provide a good model for the narrow component 
of H\al\ and \nii, then our automatic fitting procedure indeed claims to find 
a weak ($f_{blend}\,\approx$ 13\%) broad H\al\ line with FWHM $\approx$ 2200 
\kms\ (Fig. 6{\it a}).  Note, however, the sizable residuals near \nii\ 
\lamb 6583, signifying that our adopted model profile does {\it not} give a 
very good fit (the model profile is too narrow).   Furthermore, although this 
discrepancy is fairly noticeable in this instance, it could easily have been 
missed were it not for the high S/N of the spectrum.  If we instead assume that 
the profiles of narrow H\al\ and \nii\ are identical, but different 
from \sii, the algorithm reports an extremely weak broad feature ($f_{blend}
\,\approx$ 2\%), but shifted toward \nii\ \lamb 6583 (Fig. 6{\it b}; the 
broad component is not visible at the scale of the plot).  
Lacking any alternative, the latter set of conditions sometimes must be 
imposed for objects whose \sii\ profile is noticeably different from that of 
\nii\ (e.g., see NGC 4321), and in cases where the \sii\ region is simply 
too noisy (e.g., NGC 266) or corrupted (e.g., NGC 315) to yield a reliable
model profile.  Finally, making no assumptions about the similarity of the 
narrow H\al, \nii, or \sii\ profiles, {\it no} broad H\al\ component 
is required at all (Fig. 6{\it c}); the final fit finds that the FWHMs 
of H\al\ and \nii\ (both nearly identical) are only marginally (10\%)
larger than that of \sii, although their wings are significantly broader
(by about 30\%).  Thus, for NGC 1358 the most conservative conclusion, adopted 
here, is that broad H\al\ is not present.

{\it NGC 1667.} --- The narrow-line profiles have extended, non-Gaussian
wings that are slightly asymmetric to the red (Fig. 7{\it a}), giving the
false impression that there might be a broad H\al\ line in this Seyfert 2
nucleus.  Both H\al\ and \nii\ have similar profiles, but \sii\ is slightly
narrower.

{\it NGC 1961.} --- As in NGC 1667, the non-Gaussian, (blue) asymmetric wings 
of the narrow lines mimic broad H\al\ (Fig. 7{\it b}).  Close inspection of 
the high-S/N spectrum reveals that the bases of the \sii\ lines are somewhat 
narrower than those of \nii\ and H\al, whose profiles are nearly identical.  
If one had forced all the narrow lines to have the same profile as \sii, one 
would have concluded incorrectly that broad H\al\ is present {\it and} that it 
is displaced redward of the systemic velocity of the narrow component.

{\it NGC 2273.} --- Each of the narrow lines of the Seyfert 2 nucleus of 
NGC 2273 (Huchra, Wyatt, \& Davis 1982) has faint, extended wings with a 
blue tail, and they differ slightly in their profiles.  We find no evidence 
of emission from a BLR (Fig. 7{\it c}).

{\it NGC 2342.} --- Although relatively unblended, the velocity profile of each 
of the \sii\ lines requires two Gaussians for an adequate fit.  The resulting 
model roughly accounts for the H\al\ + \nii\ blend, but significant residuals 
are seen (Fig. 7{\it d}).  At any rate, broad H\al\ emission is unlikely to 
be present.

{\it NGC 2639.} --- Both Keel (1983) and Huchra \etal (1982) recognized 
that this object emits broad H\al.  The narrow-line component, however, 
must be modeled very carefully.  A detailed fit to the \sii\ lines reveals
highly irregular profiles spanning an enormous range of velocities 
(FWZI \gax 2600 \kms). These mostly account for the apparent strength of the 
broad H\al\ line, whose actual contribution is rather modest ($f_{blend}\,
\approx$ 17\%; FWHM $\approx$ 3100 \kms; Fig. 7{\it e}) and very uncertain 
(estimates differ by at least a factor of 2 among the three fitting methods).  
Previous profile decomposition of the H\al\ + \nii\ blend assuming purely 
Gaussian line shapes (e.g., Koratkar \etal 1995) has seriously overestimated 
the broad-line flux.

{\it NGC 2655.} --- The \sii\ lines yield an excellent model for the
narrow lines, and a broad H\al\ component is suggested by the decomposition 
shown on Figure 7{\it f}.  However, the result seems to be highly dependent 
on the choice of template model used in the starlight subtraction, and we
consider the detection to be questionable.

{\it NGC 2681.} --- As in NGC 3031 (Filippenko \& Sargent 1988; see also Fig. 
1{\it b}), the two \sii\ lines seem to have different widths, although the 
relatively low S/N of this wavelength region renders the fits somewhat 
uncertain.  Despite the similarity between
the profile of \nii\ and \sii\ \lamb 6731, the narrow component of H\al\
has a smaller width; the latter could be due partly to imperfect subtraction 
of H\al\ absorption, since an intermediate-age stellar template had to be 
included in the final starlight model (Paper III).  The extended tail 
redward of \nii\ \lamb 6583, apparently absent in \sii, supports the 
reality of the broad H\al\ component derived from the profile fitting, 
but its actual parameters are not well constrained (Fig. 7{\it g}).  The 
extra ``bump'' near 6523 \AA\ is due to an \ion{Fe}{1} feature which is 
stronger in one of the template galaxies (NGC 205) than in NGC 2681.

%{\it NGC 2685.} --- The low S/N spectrum of this object severely compromises .
%                 - no, maybe template mismatch!

{\it NGC 2768.} --- Although the total H\al\ + \nii\ blend hints at the 
probable presence of weak broad H\al\ wings, this suspicion is not confirmed 
by more detailed line decomposition (Fig. 7{\it h}).  This object is one 
of the rare cases in which the \sii\ lines are actually {\it broader} than 
\nii, although the magnitude of the difference is not great and needs to be 
evaluated more thoroughly.

{\it NGC 2787.} --- All of the narrow lines can be well modeled by the 
profile deduced from the \sii\ lines.  Contrary to the conclusion reached 
in Paper I, wherein the starlight was not yet removed from the spectrum, 
a fairly prominent ($f_{blend}\,\approx$ 35\%) broad H\al\ line is present 
in NGC 2787 (Fig. 8{\it a}).  The extended red wing and apparent offset of 
the central wavelength of the broad line are most likely artifacts of the 
depression in the residuals shortward of $\sim$6520 \AA.  The exact parameters 
of the broad H\al\ line in this case are very sensitive to the choice of 
templates, and we have adopted the one shown to achieve the best starlight 
subtraction over the entire spectrum.

{\it NGC 2841.} --- Despite the somewhat triangular narrow-line profiles,
a model from the \sii\ doublet gives an excellent fit to the complex (Fig. 
8{\it b}), confirming the conclusion in Paper I that broad H\al\ is absent.

{\it NGC 2911.} --- The \sii\ lines can fit the H\al\ + \nii\ blend  
closely, but not perfectly, probably because the parameters of the model 
profile are adversely affected by the somewhat noisy \sii\ region.
Nevertheless, a broad H\al\ component can probably be ruled out
(Fig. 8{\it c}).

{\it NGC 2985.} --- A moderately weak ($f_{blend}\,\approx$ 22\%) broad
H\al\ component may be necessary to account for the ``shelf'' of
excess emission discernible in the gaps between H\al\ and \nii\
(Fig. 8{\it d}).  However, because of the low quality of the data in the
\sii\ region, we cannot rule out the possibility that the excess emission
originates from weak, intrinsically broad wings in the narrow-line profile.
Thus, at this point, we will regard the broad H\al\ detection only as tentative. That the narrow component of H\al\ is slightly narrower than \nii\ (resulting
in relatively pronounced residuals) can probably be attributed to imperfect
removal of H\al\ absorption.

{\it NGC 3031.} --- As first discovered by Peimbert \& Torres-Peimbert
(1981), confirmed by Shuder \& Osterbrock (1981), and subsequently studied in
greater detail by Filippenko \& Sargent (1988), NGC 3031 (M81) has a 
conspicuous ($f_{blend}\,\approx$ 57\%) broad H\al\ line with FWHM $\approx$ 
2650 \kms.  The recent detections of nonthermal X-ray emission (Petre \etal 
1993; Ishisaki \etal 1996), a highly  compact VLBI radio core (Bietenholz 
\etal 1996), and a nonstellar, featureless ultraviolet continuum (Ho \etal 
1996) make NGC 3031 the first LINER whose multiwavelength spectrum has been 
studied in depth.  The narrow-line spectrum is characterized by a large range 
of line widths because of density stratification (Filippenko \& Sargent 1988; 
Ho \etal 1996).  Although a model of \sii\ \lamb 6731 roughly matches \nii\
and narrow H\al\ (e.g., Filippenko \& Sargent 1988, and Fig. 8{\it e}), it 
is obvious from the residuals that the base of \nii\ is somewhat wider than 
that of \sii.  Moreover, the two \sii\ lines clearly have different widths 
(Filippenko \& Sargent 1988; Fig. 1{\it b}).  The dominance of the broad H\al\ 
component makes it difficult to derive the true profile parameters of \nii.  
Both H\bet\ and H$\gamma$ have broad components (see also Filippenko \& 
Sargent 1988), and spectra taken with the {\it Hubble Space Telescope (HST)}  
show several additional broad lines in the ultraviolet (Ho \etal 1996).

{\it NGC 3079.} --- The large strengths of \oii\ \lamb 3727 (Heckman, Balick, 
\& Crane 1980) and \oi\ \lamb 6300 (Paper III) relative to \oiii\ \lamb 5007 
in the nucleus of NGC 3079 qualify it as a LINER according to the definition 
of Heckman (1980).  Adhering to the convention of Paper III, on the other 
hand, the \oiii/H\bet\ ratio puts the nucleus in the category of Seyferts; 
however, it should be noted that the H\bet\ line is very weak and 
uncertain, and the excitation could be lower.  As described extensively by 
Filippenko \& Sargent (1992) and Veilleux \etal (1994), and clearly 
illustrated in Figure 8{\it f}, the kinematics of the line-emitting gas near 
the nucleus of this edge-on SBc galaxy are extremely complicated.  The central 
starburst appears to dominate the energetics, although some contribution from 
an AGN cannot be excluded (Veilleux \etal 1994).  In view of the complexity 
of the velocity field of the gas, it is not surprising that we were 
unsuccessful in achieving a satisfactory decomposition of the H\al\ + \nii\
blend.  (We do not show any of the trial fits in Fig. 8{\it f}.)  Within our 
2\asec\ $\times$ 4\asec\ extraction aperture, the \sii\ line profile does not 
entirely match the shape of the \nii\ lines, and H\al\ seems to have a narrow 
core (Fig. 8{\it f}), perhaps due to superposed emission from \hii\ regions, 
which is absent from \nii\ and \sii.  Judging by the 
spatial variation of the emission-line spectrum (Veilleux \etal 1994), these 
profile differences most likely reflect variations in excitation or physical 
conditions among the discrete gas elements included in our integrated spectrum.
Without adequate constraints on the intrinsic narrow-line profiles, it is 
virtually impossible to fit the broad H\al\ component with confidence.  Thus, 
although Stauffer (1982) and Keel (1983) suspected broad H\al\ to be present 
in this object, we agree with Filippenko \& Sargent (1992) that it is 
premature to come to any firm conclusions.

{\it NGC 3166.} ---  As with NGC 1358, the imperfect match between the \sii\
model and the \nii\ lines can deceptively mimic a low contrast, apparently 
redshifted, broad H\al\ feature.  We will opt for the more conservative 
viewpoint that such a feature is spurious and simply fit each of the 
narrow lines with two Gaussians (Fig. 8{\it g}).

{\it NGC 3190.} --- The narrow lines are unusually broad (FWHM $\approx$ 500 
\kms) and have a clear red asymmetry, but there is no evidence for a broad 
component of H\al\ (Fig. 8{\it h}).  The H\al\ line has a noticeably 
smaller width than \nii, most likely resulting from imperfect starlight 
subtraction.

{\it NGC 3226.} --- Despite being fairly noisy, the tail redward of \nii\
\lamb 6583 is also present in \sii\ \lamb 6731.  A total of three Gaussians 
was used to model the unusual narrow-line profile, and we succeeded in 
extracting a moderately strong broad H\al\ component from the complicated 
blend (Fig. 9{\it a}).

{\it NGC 3227.} --- In order to quantitatively measure the extremely 
broad (FWZI $\approx$ 16,000 \kms) H\al\ line of this well-known Seyfert 1 
nucleus, we carefully removed the narrow-line contribution of the blend 
by scaling an accurate model constructed from \sii\ (Fig. 14{\it b}).  It is 
difficult to tell if all the narrow lines have the same profile, but
this is not crucial for the purposes of measuring such a strong, broad 
H\al\ line.  Broad H\bet\ and H$\gamma$ are both very prominent.  The 
broad lines in this object are known to be variable (e.g., Winge \etal 1995); 
the spectrum shown here was taken on 1986 March 29 UT.

{\it NGC 3245.} --- Along with NGC 5005, the line profiles of this galaxy
exhibit among the most complicated velocity structure in the survey
(Fig. 9{\it b}).  The rotation curve rises very steeply within the central 
4\asec\ in the two-dimensional spectrum, causing each 
line to have a double-horned profile.  Despite the complexity, each line can 
be well modeled by a combination of three Gaussian components: one for the 
nucleus and one each for the receding and approaching sides of the disk.  
While the same combination of components fits \nii\ and \sii, the central 
(nuclear) component of H\al\ is weaker with respect to the two disk 
components, as is obvious from the ``dip'' in the center of the profile.  This 
is simply a manifestation of the higher \nii/H\al\ and \sii/H\al\ ratios for 
the LINER component centered on the nucleus (classified in Paper III as a 
transition object).  A very faint ($f_{blend}\,\approx$ 12\%), broad 
(FWHM $\approx$ 2650 \kms) component of H\al\ is suggested by our 
decomposition, but the result, unfortunately, is ambiguous, because it seems to 
depend on the choice of template model used in the starlight subtraction.  

{\it NGC 3367.} --- This is an Sc galaxy with a prominent nucleus showing 
emission features of Wolf-Rayet stars (Paper II).  The narrow lines have 
distinctly non-Gaussian profiles with obvious blue asymmetry (Fig. 
9{\it c}).  The relatively wide bases of the narrow lines give the impression 
that there might be broad H\al, but this is not confirmed by detailed modeling 
which shows that the \sii\ lines have the same profiles.

{\it NGC 3516.} --- We determined the narrow-line spectrum of this well-known 
Seyfert 1 nucleus (Boksenberg \& Netzer 1977) quite accurately using a model 
derived from \sii\ (Fig. 14{\it c}).  The resulting fits give \nii\ 
\lamb 6583/H\al\ = 1.3, \sii\ \lamb\lamb 6716, 6731/H\al\ = 0.70, and 
\oi\ \lamb 6300/H\al\ = 0.15.  The broad-line component is also clearly 
present in H\bet\ and H$\gamma$.

{\it NGC 3642.} --- Broad H\al\ emission is easily discernible (Paper I) 
because of the small widths of the narrow lines (Fig. 9{\it d}).  As with 
many other objects, however, the broad-line component would be overestimated 
if the narrow lines were modeled as single Gaussians, since their bases have 
extended wings.  A reliable two-Gaussian model for \nii\ was derived from 
\sii; the narrow component of H\al\ evidently has a larger core-to-base 
ratio (and hence smaller FWHM) than \nii.  Since the stellar population 
of the nucleus of NGC 3642 does not appear to have exceptionally strong Balmer 
absorption, the difference in profile between H\al\ and \nii\ probably is 
not due to undersubtraction 
of H\al\ absorption; instead, the most likely
explanation is that the narrow-line spectrum is partly contaminated by
emission from circumnuclear \hii\ regions (whose \nii/H\al\ ratios are
smaller than those of AGNs).

{\it NGC 3718.} --- As noted in Paper I, broad H\al\ emission is definitely
present.  Constraining the narrow lines with a \sii\ model, we find that
a broad component with FWHM = 2350 \kms\ accounts for $\sim$50\% of the total 
flux of the blend.  The centroid of the broad component of H\al\ is redshifted
by 430 \kms\ with respect to the narrow component (Fig. 9{\it e}).  
One possible interpretation of the velocity redshift seen in the 
broad H\al\ line in NGC 3718 is that the emission is scattered off of 
outflowing electrons in the narrow-line region, as in the well-known case of 
NGC 1068 (Miller \& Antonucci 1985; Miller \etal 1991).  It should be 
remembered, however, that systematic errors in the determination of the 
continuum level or in our assumptions about the profiles of the narrow 
lines can also possibly lead to apparent velocity shifts (see notes 
on NGC 266, NGC 1358, NGC 1961, and NGC 3166).

{\it NGC 3884.} --- The extended tail of faint emission redward of \nii\
\lamb 6583 betrays a hidden broad-line component in NGC 3884 (Keel 1983; 
Paper I; Fig. 9{\it f}).  Unfortunately, the high recession velocity of the 
object shifted part of \sii\ \lamb 6731 out of our spectral range.  The 
narrow-line model profile, derived solely from \sii\ \lamb 6716, is therefore 
not well constrained.  In particular, we could not verify whether the extended 
tail of emission is truly absent from the \sii\ lines.  Nevertheless, neither 
\oi\ nor \oiii\ show any obvious signs of such extreme red asymmetry in 
their profiles, leading us to conclude that the deduced broad H\al\ component 
is very likely to be real.  

{\it NGC 3898.} --- The profiles of the \sii\ lines do not match those of    
narrow H\al\ and \nii\ in detail.  This, coupled with the fact that the 
spectrum does not have very high S/N, leads to ambiguity regarding the  
detection of broad H\al\ (Fig. 9{\it g}).  We will adopt the more 
conservative assumption that broad H\al\ is not present; it would be highly 
desirable to verify this with data of higher S/N.

{\it NGC 3982.} --- This Seyfert 2 nucleus (Paper III) contains an 
extremely faint ($f_{blend}\,\approx$ 12\%) broad H\al\ component (Fig. 
9{\it h}).  Although apparent only after detailed profile fitting, we are 
reasonably confident of this detection.  \nii\ has the same profile as 
\sii, but narrow H\al\ differs slightly.

{\it NGC 3998.} --- Broad H\al\ is unmistakably present in this famous LINER 
(Fig. 10{\it a}), as was already known from several previous 
studies (Heckman 1980; Blackman, Wilson, \& Ward 1983; Keel 1983; Paper I).  
Reichert \etal (1992) find that \mgii\ \lamb 2800 also has a broad component 
similar in width to H\al.  The narrow-line profile, as judged from the 
\sii\ doublet, is quite symmetric and can be represented by the sum of 
a narrow and a broader Gaussian.  The flux of broad H\al, as determined from 
our three standard fitting methods, showed excellent agreement (differing
by less than 5\% from each other), and we obtained the final value by
averaging all three results.  The adopted broad H\al\ component contains
$\sim$37\% of the flux of the entire H\al\ + \nii\ blend and has widths 
of FWHM $\approx$ 2150 \kms\ and FWZI \gax 5000 \kms.  Broad H\bet\ is 
also present.

{\it NGC 4036.} ---  Though faint ($f_{blend}\,\approx$ 14\%), a broad
component of H\al\ seems necessary to adequately model the H\al\ + \nii\
complex (Fig. 10{\it b}), confirming the preliminary analysis in Paper I.  The
narrow-line profile from \sii\ has extended, non-Gaussian wings, but it is
very symmetric.

{\it NGC 4051.} ---  With an absolute blue magnitude of only --15.6 
(V\'eron 1979; $H_o$ = 50 \kms\ Mpc$^{-1}$), the nucleus of NGC 4051 can be 
considered the least luminous ``classical'' type 1 Seyfert (Weedman 1976).  
As with other prominent Seyfert 1 nuclei (e.g., NGC 3516; Fig 14{\it c}), we 
removed the narrow-line contribution to the complex with a model constructed 
from the \sii\ lines (Fig. 10{\it c}).  The broad H\al\ line has unusually 
small widths (FWHM $\approx$ 1000 \kms, FWZI $\approx$ 5900 \kms) compared 
to other objects of its class (e.g., NGC 4151, NGC 5548), and this object 
most likely belongs to the group of ``narrow-lined'' Seyfert 1 nuclei 
(Osterbrock \& Pogge 1985).  Both H\bet\ and H$\gamma$ also have a broad 
component.

{\it NGC 4138.} ---  This is one of several previously unrecognized Seyfert 
galaxies discovered in our survey (Paper III).  The narrow lines 
of H\al, \nii, and \sii\ have symmetric profiles which can 
be well described by single Gaussians.  Visual inspection of the regions 
between H\al\ and \nii, further confirmed by detailed fitting, reveals a 
relatively weak broad H\al\ line responsible for $\sim$23\% of the emission 
from the entire blend (Fig. 10{\it d}).

{\it NGC 4143.} --- As the bases of the \sii\ lines are not significantly 
wider than that of a single Gaussian, the extended wings of the H\al\ + 
\nii\ complex probably signifies the presence of a prominent broad H\al\ 
component.  Indeed, a four-component Gaussian fit to the complex yields a 
broad line (FWHM $\approx$ 2100 \kms) containing a large fraction ($f_{blend}$ 
$\approx$ 45\%) of the emission (Fig. 10{\it e}).  Broad H\bet\ may also be 
present.

{\it NGC 4151.} --- The highly variable broad-line emission of this famous 
Seyfert 1 nucleus has been the target of intensive spectroscopic monitoring 
(e.g., Maoz \etal 1991; Kaspi \etal 1996).  We modeled the asymmetric 
narrow-line profile of \sii\ with a combination of three Gaussians, which were 
then scaled and subtracted to remove the narrow lines from the complex (Fig. 
14{\it d}).  The broad H\al\ profile of the spectrum shown (taken on 1986 
March 19 UT) exhibits a prominent blue asymmetry; the extra hump near 6660 
\AA\ is partially due to broad \hei\ \lamb 6678 emission.  Broad H\bet\ and 
H$\gamma$ are present in the blue spectrum, as is \heii\ \lamb 4686.

{\it NGC 4168.} ---  Because of the small equivalent widths of the emission 
lines in this elliptical galaxy, the S/N of the starlight-subtracted spectrum 
is rather low, and its classification is subject to considerable uncertainty.  
Taken at face value, the intensity ratios of the emission lines qualify 
NGC 4168 as a Seyfert galaxy (Paper III).  Despite the noise in the spectrum 
and the large uncertainties associated with the final parameters of the fit,
a rather obvious broad H\al\ feature is apparent even in the unfitted data
(Fig. 10{\it f}).

{\it NGC 4203.} --- Broad H\al\ is detected unambiguously in this LINER 
(Fig. 10{\it g}).  The fairly symmetric narrow lines of H\al, \nii, and 
\sii\ can be fitted by a single Gaussian with FWHM $\approx$ 350 \kms.

{\it NGC 4235.} --- The broad H\al\ profile of this Seyfert 1 nucleus 
(Abell, Eastmond, \& Jenner 1978) has a peculiar secondary hump redward of 
\nii\ \lamb 6583 resembling that in NGC 7213 (Filippenko \& Halpern 1984).  
The narrow-line profile, as judged by modeling \sii, requires a 
superposition of three Gaussians to reproduce its irregular shape (Fig. 
14{\it e}).  The profile of \nii\ \lamb 6583 appears to have additional 
structure not present in \sii, but, for the present purposes, we neglect this 
subtlety.  A very prominent broad component is visible at H\bet.

{\it NGC 4258.} --- With \oiii\ \lamb 5007/H\bet\ $\approx$ 10, NGC 4258 (M106)
harbors a Seyfert nucleus (Paper III), not a LINER (e.g., Heckman 1980;
Greenhill \etal 1995).  VLBI observations recently have furnished dynamical 
evidence that the nucleus has an extremely high mass density, probably
indicative of a supermassive black hole (Miyoshi \etal 1995; Greenhill \etal 
1995).  A broad H\al\ component is apparent in the high-resolution nuclear 
spectra published by St\"uwe, Schulz, \& H\"uhnermann (1992; see their Fig. 
2), although the detailed shape of the H\al\ + \nii\ complex is somewhat 
different from ours (Fig. 10{\it h}) because these authors did not correct 
their data for starlight contamination.  St\"uwe et al. questioned the reality 
of the broad H\al\ line, arguing that Lorentzian profiles alone, instead of 
Gaussians, can give an acceptable fit to the complex without invoking an extra 
broad component.  But as we have shown throughout this paper, any single 
analytic profile, whatever its form, generally gives a poor representation of
the intrinsic shapes of the narrow lines, and a more involved procedure must 
be applied.  Using a model for the narrow lines constrained by the \sii\ 
profile, a very pronounced broad H\al\ component, as was previously noticed by 
Stauffer (1982) and Paper I, is inferred from our fit (Fig. 10{\it h}).  The 
narrow lines have an obvious blue asymmetry and FWHMs of about 
290 \kms.  The broad H\al\ component can be represented reasonably well by a 
Gaussian of FWHM $\approx$ 1700 \kms, from which we deduce a luminosity (not 
corrected for extinction) of 4.5\e{38} \lum\ (assuming a distance of 6.8 Mpc; 
Tully 1988).  Note that Wilkes \etal (1996) found that the polarized 
spectrum of the nucleus exhibits emission lines that are broader 
than those in the unpolarized spectrum.  The ``broad'' lines in this case, 
however, only have FWHM $\approx$ 1000 \kms\ and are seen both in the 
permitted lines as well as the forbidden lines.  This component, therefore, 
is different from the broad component we observe associated with H\al, which 
we attribute to the classical BLR.  A weak broad H\al\ component may be 
present in the polarized spectrum of Wilkes et al., but the S/N of their data 
is insufficient to tell.

{\it NGC 4261.} --- Images from the {\it HST} reveal an extended disk 
surrounding the LINER nucleus of NGC 4261 (Jaffe \etal 1993), and dynamical 
analysis suggests the presence of a supermassive black hole (Jaffe \etal 1996; 
Ferrarese, Ford, \& Jaffe 1996).  The \sii\ lines in our data are not of 
sufficiently high quality to yield a reliable model for the narrow lines.  
Nevertheless, all the narrow lines look quite symmetric and have broad wings.  
We represented each line with a combination of two Gaussians centered at the 
same wavelength (Fig. 11{\it a}).  In most of the lines (H$\gamma$, 
H\bet, \nii, H\al, and \sii), the narrow core has FWHM $\approx$ 300 
\kms, and the broader base has FWHM $\approx$ 900 \kms.  Some of the forbidden 
lines with higher critical densities, however, also have larger line widths.  
\oiii\ \lamb\lamb 4959, 5007 has a core with FWHM $\approx$ 300 \kms\ but a 
base with FWHM $\approx$ 1500 \kms, while \oi\ \lamb 6300 is best modeled 
with two Gaussians of FWHM $\approx$ 650 \kms\ and 1400 \kms.  Jaffe \etal 
(1996) recently performed a detailed analysis of the optical emission-line 
profiles of the nucleus of NGC 4261 based on ground-based spectra.  Their 
results are consistent with ours, the most noticeable difference being 
that they detected much broader lines than we did.  The \nii, H\al, and \sii\ 
lines in the Jaffe et al. spectrum, for instance, have cores with FWHM 
$\approx$ 410--490 \kms\ and wings with FWHM $\approx$ 2700--3500 \kms.  
We attribute this difference to the fact that Jaffe et al. acquired their 
data using a narrower slit than we did (1\farcs1 versus 2\asec) and 
under conditions of better seeing (1\asec\ versus 1\farcs5).  Because the 
widths of the emission lines in NGC 4261 increase very rapidly close to the 
nucleus (Ferrarese \etal 1996), the better spatial resolution of the 
Jaffe et al. spectrum naturally samples gas with higher velocities.
Consistent with Jaffe et al., we find no H\al\ emission arising from a BLR.
The small-aperture (0\farcs1) {\it HST} spectra of Ferrarese \etal (1996) also 
show no obvious signs of broad H\al.

{\it NGC 4278.} ---  As illustrated in Figure 2, a relatively weak, broad
H\al\ line can be extracted from NGC 4278, an elliptical radio galaxy with a
LINER nucleus.  Rotational broadening contributes to the large width
(FWHM = 480 \kms) of the narrow lines, each of which is well modeled by the
sum of two Gaussians.

{\it NGC 4321.} --- The extended and highly asymmetric wings of the narrow 
lines in the transition nucleus of NGC 4321 (M100) superficially resemble broad 
H\al\ emission.  Indeed, if one naively assumed that the \sii\ lines can 
be used to model the narrow component of H\al\ and \nii, one could force 
a broad feature in the fit (e.g., as in NGC 1358).  Fortunately, the data 
have high enough S/N that differences in the narrow-line profiles are 
discernible.  An excellent fit to the complex can be achieved simply by 
constraining narrow H\al\ and \nii\ to have identical profiles (Fig. 
11{\it b}), and no broad H\al\ is necessary for this object.

{\it NGC 4388.} --- Originally classified as a Seyfert 2 galaxy (Phillips 
\& Malin 1982), the nucleus of NGC 4388 in fact contains very faint 
wings of broad H\al\ emission (Stauffer 1982; Paper I) with FWZI $\approx$ 
6000 \kms\ (Shields \& Filippenko 1988, 1996).  In addition, Shields \& 
Filippenko discovered broad H\al\ emission in at least two off-nuclear 
positions, which they interpreted as scattered radiation from a largely hidden 
Seyfert 1 nucleus, similar to that seen in NGC 1068 (Miller \etal 1991).
As in most nuclei discussed in this study, the narrow lines 
themselves have extended, asymmetric wings; these have to be properly removed 
to accurately measure the broad H\al\ flux (Fig. 11{\it c}).  Using a model 
profile derived from \sii, each narrow line is well represented by a 
combination of three Gaussians.  The resulting broad line (FWHM $\approx$ 3900 
\kms) contains a mere 2$\pm$2\% of the flux of the entire blend.  Because of 
the scale of the plot, it is barely visible in Figure 11, but Figure 14 in 
Paper I more clearly illustrates its presence.

{\it NGC 4395.} --- This galaxy holds the special distinction of hosting the 
least luminous ($M_B$ = --9.8 mag) and most nearby (2.6 Mpc) Seyfert 1 nucleus 
known (Filippenko \& Sargent 1989).  As in Filippenko \& Sargent, the 
narrow components of H\al\ and \nii\ were removed from the blend using \sii\
as a model (Fig. 11{\it d}).  The narrow lines, whose FWHMs are 40--50 \kms\ 
(Filippenko \& Ho 1997), are unresolved at the current resolution.  As 
confirmed by the high-resolution spectra of Filippenko \& Ho, the narrow-line 
profiles all have a blue asymmetric tail.  Although the FWHM of the broad 
H\al\ component is only $\sim$500 \kms, the base of the line stretches to 
a FWZI of $\sim$5000 \kms\ (less than the 7000 \kms\ quoted by Filippenko \& 
Sargent 1989).  A broad component is present in a number of other 
optical (Filippenko \& Sargent 1989) and ultraviolet (Filippenko, Ho, \& 
Sargent 1993) emission lines.

{\it NGC 4438.} --- With a strength of only 9\% of the total H\al\ + \nii\
blend, the broad H\al\ component in NGC 4438 is among the weakest of our 
probable detections.  The high S/N of the starlight-subtracted spectrum 
allowed us to construct an accurate profile model from \sii, which matched 
well the profiles of the other narrow lines.  Very faint residual wings were 
evident on either side of \nii\ even after the narrow lines were removed, and 
the best fit was achieved with the addition of a broad Gaussian with 
FWHM $\approx$ 2050 \kms\ (Fig. 11{\it e}).

{\it NGC 4450.} ---  The narrow-line profiles of this LINER show extended, 
slightly asymmetric wings which largely contribute to the broad base of the 
complex (Fig. 11{\it f}).  However, additional low-level emission extends 
beyond the wings of \nii, and a full analysis of the blend indicates the 
probable presence of a fairly weak ($f_{blend}$ $\approx$ 20\%; FWHM $\approx$ 
2300 \kms) broad H\al\ line, confirming the suggestions of Stauffer (1982) 
and Paper I.

{\it NGC 4486.} ---  Although the presence of weak extensions on either side 
of the \nii\ lines led us to suspect in Paper I that broad H\al\ might exist 
in the nucleus of this famous active galaxy (M87), careful inspection of the 
spectrum after starlight subtraction indicates that the \sii\ lines may have 
similar wings in their profiles.  Note also that the nuclear spectrum obtained 
with the {\it HST} using a 0\farcs26 aperture (Harms \etal 1995) shows that 
both \nii\ and \sii\ have similar profiles.  The enormous widths of the 
emission lines (FWHM $\approx$ 1000--1100 \kms), unfortunately, causes severe 
blending, as a consequence of which a unique model cannot be determined for 
the highly asymmetric \sii\ lines based on our data.  With the assumption 
that both \sii\ lines have the same profile, each line can be modeled by a 
combination of four Gaussian components.  If we apply this model to all three 
lines in the H\al\ + \nii\ blend, it can be seen that the fit is rather poor 
(Fig. 11{\it g}).  Adding an additional broad Gaussian to the fit will 
obviously reduce the residuals, and such a component can be interpreted as
broad H\al\ emission, but we choose not to do so because of the above 
ambiguities.  The small-aperture spectrum of Harms et al. also does not 
show broad emission associated with H\al\ apart from the extended wings seen 
in all the narrow lines.

{\it NGC 4501.} ---  NGC 4501 harbors a previously unrecognized Seyfert 2
nucleus (Paper III).  Judging by the faint wings on either side of \nii, it
appears that an extremely faint ($f_{blend}$ $\approx$ 9\%), broad (FWZI
$\approx$ 5000 \kms) H\al\ component is present (Fig. 11{\it h}), but its
detailed parameters are uncertain as a result of ambiguities in the true shape 
of the narrow lines.
 
{\it NGC 4565.} --- The large \oiii/H\bet\ ($\sim$9) and \nii/H\al\
($\sim$2.5) ratios qualify NGC 4565 as a Seyfert galaxy.  A relatively faint
($f_{blend}$ $\approx$ 16\%) broad H\al\ component is seen in the H\al\ +
\nii\ blend after fitting the narrow lines with a model of \sii\ \lamb 6731
(Fig. 12{\it a}).  If this detection is correct, the broad H\al\ luminosity 
(8\e{37} \lum) is the lowest yet found in any active nucleus, being even 
smaller than that of NGC 4395.  As in the case of NGC 3031 (Fig. 1{\it b}), 
\sii\ \lamb 6716 is narrower than \sii\ \lamb 6731.
 
{\it NGC 4579.} ---  Broad H\al\ emission has long been known to be present 
in this LINER/Seyfert nucleus (Stauffer 1982; Keel 1983; Paper I), but its 
strength is substantially weaker than that deduced from decomposition of the 
narrow lines using single Gaussians.  The extended, asymmetric base of \sii\ 
largely accounts for most of the broad base of the H\al\ + \nii\ complex.  
Nevertheless, a detailed multi-component fit undeniably requires a moderately 
strong ($f_{blend}$ $\approx$ 21\%) broad H\al\ line (Fig. 12{\it b}).  The 
BLR in NGC 4579 recently has also been revealed in the ultraviolet (Barth 
\etal 1996).

{\it NGC 4594.} --- Contrary to the conclusion in Paper I, we find no evidence 
of broad H\al\ in the LINER nucleus of NGC 4594 (the Sombrero galaxy; M104). 
Careful inspection of the line profiles indicates that the \sii\ lines have 
large widths (FWHM $\approx$ 500 \kms) and extended wings (FWZI $\approx$ 3000 
\kms); in fact, this is one of the rare instances in which the \sii\ lines are 
actually broader than the narrow H\al\ and \nii\ lines.  The blend was 
modeled assuming that H\al\ and \nii\ have identical profiles, each 
of which can be represented by the sum of two Gaussians (Fig. 12{\it c}).
Note that broad H\al\ {\it has} been detected in the recent, small-aperture 
{\it HST} spectra of Kormendy \etal (1997), illustrating the advantages that 
can be gained from data having high spatial resolution.

{\it NGC 4636.} --- A moderately strong broad H\al\ component is apparent in 
the starlight-subtracted spectrum (Fig. 12{\it d}).  Although the \sii\
region is too noisy to provide a reliable model for the narrow lines, the 
narrow H\al\ and \nii\ lines are sufficiently Gaussian in shape that a simple 
four-component fit is adequate to isolate broad H\al.  As with NGC 3031 and 
NGC 4565, \sii\ \lamb 6716 is narrower than \sii\ \lamb 6731, indicative of 
density stratification in the narrow-line region (Filippenko \& Sargent 1988).

{\it NGC 4639.} --- This rather prominent Seyfert 1 galaxy had gone unnoticed 
prior to our survey (Filippenko \& Sargent 1986).  To better characterize 
the broad H\al\ component, we removed the narrow-line contribution using 
a \sii\ model (Fig. 14{\it f}).  The final broad profile (observed on 1985 
February 24 UT) has FWHM $\approx$ 3600 \kms\ and FWZI $\approx$ 9600 \kms; 
a separate spectrum obtained about a year later showed flux variability at the 
level of $\sim$10\%.  A broad component is seen in H\bet, and possibly also 
in H$\gamma$.

{\it NGC 4651.} ---  Although we suspected that broad H\al\ might be present 
at a very faint level, we could not confirm its reality unambiguously.  In 
particular, the line profiles of H\al, \nii, and \sii\ do not agree in 
detail, perhaps a consequence of the difficulty encountered during the 
starlight subtraction for this object.  Each of the narrow lines of the 
H\al\ + \nii\ blend was simply fitted with two Gaussians (Fig. 12{\it e}).

{\it NGC 4698.} ---  The emission lines of this low-luminosity Seyfert nucleus 
are quite narrow (FWHM $\approx$ 170 \kms), despite the fairly early Hubble 
type (Sa) of the host galaxy.  No trace of broad H\al\ is visible in the 
relatively high S/N spectrum (Fig. 12{\it f}).

{\it NGC 4750.} --- Broad H\al\ emission definitely exists in this LINER.  The 
high S/N spectrum reveals faint wings extending well beyond the intrinsic 
profile of \nii\ (Fig. 12{\it g}).  The narrow-line profile obtained from 
\sii\ looks very symmetric.

{\it NGC 4772.} --- Broad H\al\ is present unambiguously, as is evident from 
the wide base on either side of the \nii\ lines (Fig. 12{\it h}).  
Unfortunately, undulations in the neighboring continuum compromise the 
accuracy of the fit to the broad feature.  Because of the steep velocity 
gradient in the inner rotation curve, the narrow lines have a double-peaked or 
flat-topped profile, which was well reproduced using two Gaussian components 
derived by fitting the \sii\ lines.  Although the fit to \nii\ \lamb 6583 
shows considerable residuals, we chose to impose stringent constraints because 
of the complexity of the blend.

{\it NGC 5005.} --- Despite the extremely complicated velocity structure 
and severe blending of the narrow lines, once again caused by a steep gradient 
in the inner rotation curve, a rather significant ($f_{blend}$ $\approx$ 33\%; 
Fig. 13{\it a}) broad H\al\ component is suggested if one requires that narrow 
H\al\ and \nii\ have the same profiles as \sii.  The sizable residuals
notwithstanding, it is remarkable that such simple constraints can reproduce
most of the features of the complex blend.

{\it NGC 5033.} --- The nucleus of NGC 5033 has been detected in X-rays
(Halpern \& Steiner 1983), and the emission comes from an unresolved source
(Koratkar \etal 1995).  The very prominent and variable (Paper I) broad H\al\
line, originally noted by Shuder (1980) and Stauffer (1982), can be fitted
fairly accurately with a Gaussian of FWHM $\approx$ 2850 \kms\ (Fig.
13{\it b}).  The narrow components of H\al\ and \nii, all of which show
asymmetric wings, were modeled using the \sii\ lines.  H\bet\ also has a 
visible broad component.

{\it NGC 5077.} --- The derived strength of the broad H\al\ component in this 
object (Paper I) depends critically on the assumption about the intrinsic \sii\
profile, which, unfortunately, is rather uncertain with the present S/N of 
the spectrum.  In particular, different (equally acceptable) assumptions about
the \sii\ profile can lead to a factor of 3 discrepancy in the final broad 
H\al\ flux.  In Figure 13{\it c}, we have chosen the set of parameters which 
gives the smallest probable broad H\al\ flux.

{\it NGC 5194.} --- The broad base of the H\al\ + \nii\ blend in NGC 5194 
(M51) is predominantly due to the wide, asymmetric wings of the narrow lines. 
But careful scrutiny reveals that the extremes of the 
\nii\ lines may extend to a FWZI $\approx$ 6000 \kms\ at the faintest flux 
levels (Fig. 13{\it d}).  Although we cannot formally recover such a weak 
feature with our line-fitting technique, it is possible that broad H\al\ 
is present.  

{\it NGC 5273.} --- The broad and variable (Stauffer 1982; Paper I) H\al\ 
emission of the low-luminosity Seyfert 1 nucleus in NGC 5273 (Fig. 14{\it g}) 
closely resembles that of NGC 4639 (Fig. 14{\it f}) in relative strength 
($f_{blend}$ $\approx$ 84\%) and profile (FWHM $\approx$ 3350 \kms; FWZI 
$\approx$ 10,000 \kms).  As in NGC 4639, soft X-ray emission has been detected 
from the compact nucleus (Koratkar \etal 1995).  The \nii\ lines and the 
narrow component of H\al\ are not well fitted by the model profile derived from 
\sii, although in this instance the mismatch has an insignificant effect 
on the measured broad H\al\ flux.  Both H\bet\ and H$\gamma$ also have a 
broad component.

{\it NGC 5548.} --- This famous Seyfert 1 galaxy has been the target of 
recent intensive variability monitoring aimed at determining the size and 
structure of its BLR (see Peterson 1993 for a review).  The 
\sii\ lines have FWHM $\approx$ 300 \kms\ and display an obvious blue wing.  
Constraining the narrow H\al\ and \nii\ lines to have the same profiles, we 
subtracted suitably scaled versions of the \sii\ profile from the H\al\ + 
\nii\ complex (Fig. 14{\it h}).  The final strengths of the narrow lines, 
determined such that the broad profile be as smooth as possible at their 
positions, yield the following narrow-line intensity ratios: 
\nii\ \lamb 6583/H\al\ = 0.9, \sii\ \lamb\lamb 6716, 6731/H\al\ = 0.7, 
\oi\ \lamb 6300/H\al\ = 0.4, and \oiii\ \lamb 5007/H\bet\ = 10. In addition to 
the narrow and broad components of H\al, there is clearly also substantial 
emission from a distinct component with intermediate velocities (FWHM 
$\approx$ 1000 \kms).  Although the intermediate-width component is much less 
prominent in H\bet, comparison of the profile of \oiii\ \lamb 5007 with that of 
the central peak of H\bet\ suggests that it might also be present.  Broad 
H\bet\ and H$\gamma$ are present in the blue spectrum, as is an \ion{Fe}{2} 
complex centered near 4550 \AA, and possibly also \heii\ \lamb 4686.

{\it NGC 6500.} --- The spectrum of NGC 6500 looks remarkably similar to that 
of NGC 1052, both in terms of the widths and the relative strengths of the 
emission lines.  But unlike the case of NGC 1052, our profile decomposition 
suggests that broad H\al\ emission is absent in NGC 6500 (Fig. 13{\it e}).  A 
simple two-Gaussian model from \sii\ gives an excellent fit to H\al\ and \nii, 
and the extensions on either side of the blend clearly belong to the intrinsic 
profile of \nii.  The morphology of the radio continuum emission of NGC 6500 
has been interpreted as evidence for a bipolar outflow along the minor axis 
of the galaxy (Unger, Pedlar, \& Hummel 1989), similar in nature to the 
wind-driven bubbles seen in M82 and NGC 3079 (Heckman, Armus, \& Miley 
1990; Filippenko \& Sargent 1992).  The large line widths of the nucleus of 
NGC 6500 (FWHM $\approx$ 500--600 \kms), however, do not seem to be a direct 
consequence of the outflow (if indeed present); instead, the two-dimensional 
spectrum indicates that a steeply rising rotation curve is responsible for the 
line broadening.

{\it NGC 6951.} --- No broad H\al\ line is found in the Seyfert nucleus of 
this galaxy (Fig. 13{\it f}).  The emission lines are slightly asymmetric but 
well separated at our resolution, and a simple two-component model for the 
narrow-line profile gives a satisfactory fit to the spectrum.

{\it NGC 7217.} --- As with NGC 5194, the narrow lines have wide wings that 
account for most of the broad base of the H\al\ + \nii\ blend (Fig. 
13{\it g}).  However, an extremely weak component of broad H\al\ may be 
present, as suggested by the very faint wings on either end of \nii.  
Spectra having higher S/N are necessary to confirm this.

{\it NGC 7479.} --- The velocity structure of NGC 7479 (Fig. 13{\it h}) is 
reminiscent of that of NGC 3245 (Fig. 9{\it b}), except that in this case 
the narrow lines have a prominent blue asymmetry.  Note that H\al\ and  \nii\
have slightly different profiles, reflecting differences in the \nii/H\al\ 
ratio between the two main velocity components of the line.  Extremely faint 
wings, amounting to only $\sim$8\% of the flux of the H\al\ + \nii\ blend, 
can be seen with FWZI $\approx$ 3700 \kms.

{\it Special Cases.} --- A broad base (FWHM $\approx$ 1500--2000 \kms) 
sometimes accompanies the H\al\ line in galaxies whose spectrum is dominated by 
emission from Wolf-Rayet stars (so-called Wolf-Rayet galaxies).  One such 
object, discussed by Sargent \& Filippenko (1991), is NGC 4214, and several 
other examples found in our survey (IC 10, NGC 1156, NGC 1569, NGC 4532) were 
noted in Paper II.  We do not include these cases into our overall statistics 
of broad H\al\ emission (\S\ 4), since in this paper we are only concerned 
with emission arising from AGNs.

\section{Survey Statistics and Conclusions}

An optical spectroscopic survey specifically designed to detect low-luminosity 
AGNs has recently been completed.  This paper presents the data for the subset 
of nuclei showing definite or probable evidence of broad H\al\ emission --- 
similar in character but much weaker in strength than that of more luminous 
type 1 Seyferts.  The overall statistics of the survey can be summarized as 
follows: of the 211 emission-line nuclei classified as LINERs, transition 
objects (composite \hii\ nucleus/LINER), or Seyferts (see Paper III for 
classification of all objects), 34 (16\%) definitely have broad H\al, and an
additional 12 (6\%) probably do.  Questionable detections were found in 
another 18 objects (Table 2), while 27 objects suspected to have broad 
H\al\ based on visual inspection turned out to yield negative results 
(Table 3).  Thus, approximately 20\% of all nearby AGNs, corresponding to 
$\sim$10\% of all nearby, bright ($B_T\,\leq$ 12.5 mag) galaxies, can be 
considered to be AGNs harboring a BLR (``type 1'' objects).  Of 
the 34 objects with definite detections of broad H\al, only 9 are well-known 
Seyfert 1 galaxies (V\'eron-Cetty \& V\'eron 1996).  The majority have 
substantially lower H\al\ luminosities, and many of the detections or probable 
detections described in this paper are reported for the first time.  Although 
the presence of broad H\al\ emission was previously suspected in a number 
of objects, quantitative measurements were lacking.  As demonstrated in 
\S\ 3, in most instances the detection of broad H\al\ and subsequent 
determination of its line parameters require data of extremely high quality, 
sufficient spectral resolution, careful treatment of starlight contamination, 
and meticulous profile decomposition.

Excluding all the previously recognized (V\'eron-Cetty \& V\'eron 1996) 
Seyfert 1 nuclei (retaining only NGC 4395), the broad H\al\ lines of 
the remaining objects have typical luminosities of $\sim$10$^{39}$ 
\lum, FWHM $\approx$ 2200 \kms, and constitute 10\%--50\% of the H\al\ + 
\nii\ blend.  Five sources have broad H\al\ luminosities of only 
(1--3)\e{38} \lum. The lowest luminosity source with a probable detection is 
the nucleus of NGC 4565, having a broad H\al\ luminosity of only 8\e{37} 
\lum.  The luminosity function of the sources described in this study will 
be considered in a forthcoming paper.

Of the 46 objects with broad H\al\ emission, more than half belong to the 
LINER category (22 LINERs and 2 transition objects).  In the context of 
trying to decipher the physical origin of this class of objects and to 
establish their relationship to Seyfert nuclei, this is an 
important finding.  It implies that LINERs, like Seyferts, evidently come 
in two flavors --- some have a visible BLR, and others do not.
By direct analogy with the nomenclature established for Seyferts, we 
proposed in Paper III that the ``type 1'' and ``type 2'' designations be 
extended to include LINERs and transition objects.  

The broad H\al\ detection rates reported here represent lower limits to the 
true incidence of broad-line emitting regions in the sample objects.  It 
should be clear from the examples given in \S\ 3 that ambiguities concerning 
the reality of the line are encountered in some instances, and, from our 
discussion of the likely systematic errors (\S\ 2.4), there is a practical 
limit to which a weak, broad spectral feature can be extracted from the 
integrated spectra.  Detecting the broad line may be especially challenging 
in AGNs whose integrated spectrum is heavily contaminated by emission from 
nearby \hii\ regions.  In such cases, we anticipate the AGN component, 
including any possible BLR emission, to be substantially diluted.  This 
may explain why broad H\al\ was detected in so few of the transition 
objects (only 2 out of 65 sources; see Paper V) compared to the LINERs 
(22 out of 94 sources) or the Seyferts (22 out of 52 sources).
Thus, although our survey for broad H\al\ emission is 
the most comprehensive and sensitive search of its kind, it should not be 
regarded as ``complete.''  To significantly improve the sensitivity 
to even weaker emission would require observations at much higher spatial 
resolution in order to better isolate the nucleus from the contaminating bulge 
starlight.  The superior image quality of the {\it HST}, especially in 
conjunction with the use of small apertures, should be especially useful 
in this regard.  As a concrete illustration of the advantages to be gained, 
we note that Kormendy \etal (1997) discovered a broad component to H\al\ 
in their recent {\it HST} spectra of NGC 4594, whereas we could not detect 
it in our ground-based data (\S\ 3).  Thus, a concerted spectroscopic 
follow-up program using the {\it HST} would be of considerable value to 
ascertain the completeness of our ground-based survey.

\acknowledgments

The research of L.~C.~H. is currently funded by a postdoctoral fellowship
from the Harvard-Smithsonian Center for Astrophysics.  Financial support for
this work was provided by NSF grants AST-8957063 and AST-9221365, as well as
by NASA grants AR-5291-93A and AR-5792-94A from the Space Telescope Science
Institute (operated by AURA, Inc., under NASA contract NAS5-26555).
We thank Chris McKee and Hy Spinrad for their critical reading of an earlier 
draft of the manuscript, and an anonymous referee for suggesting some 
points of clarification.
%We made use of the NASA/IPAC Extragalactic Database (NED) which is operated 
%by the Jet Propulsion Laboratory, California Institute of Technology, under 
%contract with the National Aeronautics and Space Administration

%REFERENCES
\clearpage

\centerline{\bf{References}}
\medskip

\refindent
Abell, G.~O., Eastmond, T.~S., \& Jenner, D.~C. 1978, \apj, 221, L1

\refindent
Alloin, D., Pelat, D., Boksenberg, A., \& Sargent, W.~L.~W. 1983, \apj, 275,
493

\refindent
Antonucci, R.~R.~J., \& Miller, J.~S. 1985, \apj, 297, 621

\refindent
Bahcall, J.~N., Kirhakos, S., Saxe, D.~H., \& Schneider, D.~P. 1997, \apj, in
press

\refindent
Baldwin, J.~A., Wilson, A.~S., \& Whittle, M. 1987, \apj, 319, 84

\refindent
Balick, B., \& Heckman, T.~M. 1983, \apj, 265, L1

\refindent
Barth, A.~J., Reichert, G.~A., Filippenko, A.~V., Ho, L.~C., Shields, J.~C., 
Mushotzky, R.~F., \& Puchnarewicz, E.~M. 1996, \aj, 112, 1829

\refindent
Bietenholz, M., \etal 1996, \apj, 457, 604

\refindent
Blackman, C.~P., Wilson, A.~S., \& Ward, M.~J. 1983, \mnras, 202, 1001

\refindent
Boksenberg, A., \& Netzer, H. 1977, \apj, 212, 37

\refindent
Boroson, T.~A., Oke, J.~B., \& Green, R.~F. 1982, \apj, 263, 32

\refindent
Boyce, P.~J., Disney, M.~J., Blades, J.~C., Boksenberg, A., Crane, P.,
Deharveng, J.~M., Macchetto, F.~D., Mackay, C.~D., \& Sparks, W.~B. 1996,
\apj, 473, 760

\refindent
Caganoff, S., \etal 1991, \apj, 377, L9

\refindent
Cecil, G., Bland, J., \& Tully, R.~B. 1990, \apj, 355, 70

\refindent
De Robertis, M.~M., \& Osterbrock, D.~E. 1986, \apj, 301, 727

%\refindent
%de Vaucouleurs, G., de Vaucouleurs, A., Corwin, H.~G., Jr., Buta, R.~J.,
%Paturel, G., \& Fouqu\'e, R. 1991, Third Reference Catalogue of Bright
%Galaxies (New York: Springer)

\refindent
Ferrarese, L., Ford, H.~C., \& Jaffe, W. 1996, \apj, 470, 444

\refindent
Filippenko, A.~V. 1985, \apj, 289, 475

%\refindent
%Filippenko, A. V. 1987, in Observational Evidence of Activity in Galaxies,
%ed. E. Ye. Khachikian, K.~J. Fricke, \& J. Melnick (Dordrecht: Reidel), 451

\refindent
Filippenko, A. V. 1989, in Active Galactic Nuclei, ed. D.~E. Osterbrock \&
J.~S. Miller (Dordrecht: Kluwer), 495

%\refindent
%Filippenko, A. V. 1993, in The Nearest Active Galaxies, ed. J. Beckman,
%L. Colina, \& H. Netzer (Madrid: CSIC Press), 99

\refindent
Filippenko, A.~V. 1996, in The Physics of LINERs in View of Recent
Observations, ed.  M. Eracleous, et al. (San Francisco: ASP), 17

\refindent
Filippenko, A.V., \& Halpern, J.~P. 1984, \apj, 285, 458

\refindent
Filippenko, A.~V., \& Ho, L.~C. 1997, \apj, submitted

\refindent
Filippenko, A.~V., Ho, L.~C., \& Sargent, W.~L.~W. 1993, \apj, 410, L75

\refindent
Filippenko, A.~V., \& Sargent, W.~L.~W. 1985, \apjs, 57, 503 (Paper I)

\refindent
Filippenko, A.~V., \& Sargent, W.~L.~W. 1986, in Structure and Evolution
of Active Galactic Nuclei, ed. G. Giuricin \etal (Dordrecht: Reidel), 21

\refindent
Filippenko, A.~V., \& Sargent, W.~L.~W. 1988, \apj, 324, 134

\refindent
Filippenko, A.~V., \& Sargent, W.~L.~W. 1989, \apj, 342, L11

\refindent
Filippenko, A.~V., \& Sargent, W.~L.~W. 1992, \aj, 103, 28

\refindent
Fosbury, R.~A.~E., Melbold, U., Goss, W.~M., \& Dopita, M.~A. 1978,
\mnras, 183, 549

\refindent
Giovannini, G., Feretti, L., \& Comoretto, G. 1989, \apj, 358, 159

\refindent
Greenhill, L.~J., Jiang, D.~R., Moran, J.~M., Reid, M.~J., Lo, K.-Y., \&
Claussen, M.~J. 1995, \apj, 440, 619

\refindent
Halpern, J.~P., \& Steiner, J.~E. 1983, \apj, 269, L37

\refindent
Harms, R.~J., \etal 1995, \apj, 435, L35

\refindent
Heckman, T.~M. 1980, \aa, 87, 152

\refindent
Heckman, T.~M., Armus, L., \& Miley, G.~K. 1990, \apjs, 74, 833

\refindent
Heckman, T.~M., Balick, B., \& Crane, P.~C. 1980, \aas, 40, 295

\refindent
Ho, L.~C. 1995, Ph.D. thesis, Univ. of California at Berkeley

\refindent
Ho, L.~C. 1996, in The Physics of LINERs in View of Recent Observations, ed.
M. Eracleous, et al. (San Francisco: ASP), 103

\refindent
Ho, L.~C., Filippenko, A.~V., \& Sargent, W.~L.~W. 1993, \apj, 417, 63

\refindent
Ho, L.~C., Filippenko, A.~V., \& Sargent, W.~L.~W. 1994, in IAU Symp. 159,
Multi-Wavelength Continuum Emission of AGN, ed.  T.~J.-L. Courvoisier \& A.
Blecha (Dordrecht: Reidel), 275

\refindent
Ho, L.~C., Filippenko, A.~V., \& Sargent, W.~L.~W. 1995, \apjs, 98, 477 
(Paper II)

\refindent
Ho, L.~C., Filippenko, A.~V., \& Sargent, W.~L.~W. 1996, \apj, 462, 183
 
\refindent
Ho, L.~C., Filippenko, A.~V., \& Sargent, W.~L.~W. 1997a, \apjs, in press
(Paper III)
 
\refindent
Ho, L.~C., Filippenko, A.~V., \& Sargent, W.~L.~W. 1997b, \apj, in press
(Paper V)
 
\refindent
Huchra, J.~P., Wyatt, W.~F., \& Davis, M. 1982, \aj, 87, 1628

\refindent
Ishisaki, Y., \etal 1996, PASJ, 48, 237

\refindent
Jaffe, W., Ford, H.~C., Ferrarese, L., van den Bosch, F., \& O'Connell, R.~W.
1993, \nat, 364, 213

\refindent
Jaffe, W., Ford, H.~C., Ferrarese, L., van den Bosch, F., \& O'Connell, R.~W.
1996, \apj, 460, 214

\refindent
Kaspi, S., \etal 1996, \apj, 470, 336

\refindent
Keel, W.~C. 1983, \apj, 269, 466

\refindent
Keel, W.~C., \& Miller, J.~S. 1983, \apj, 266, L89

\refindent
Khachikian, E.~Y., \& Weedman, D.~W. 1974, \apj, 192, 581

\refindent
Koratkar, A.~P., Deustua, S., Heckman, T.~M., Filippenko, A.~V., Ho, L.~C., \&
Rao, M. 1995, \apj, 440, 132

\refindent
Kormendy, J., \etal 1997, \apj, 473, L91

\refindent
Malkan, M.~A., \& Filippenko, A.~V. 1983, \apj, 275, 477

\refindent
Malkan, M.~A., Margon, B., Chanan, G.~A. 1984, \apj, 280, 66

\refindent
Maoz, D., \etal 1991, \apj, 367, 493

\refindent
McLeod, K.~K., \& Rieke, G.~H. 1994a, \apj, 420, 58

\refindent
McLeod, K.~K., \& Rieke, G.~H. 1994b, \apj, 431, 137

\refindent
Miller, J.~S., Goodrich, R.~W., \& Mathews, W.~G. 1991, \apj, 378, 47

\refindent
Miller, J.~S., Tran, H., \& Sheinis, A. 1996, \baas, 28, 1301

\refindent
Miyoshi, M., \etal 1995, \nat, 373, 127

\refindent
Netzer, H. 1990, in Active Galactic Nuclei, SAAS-FEE Advanced Course 20,
Swiss Society for Astrophysics and Astronomy, ed. T.~J.-L. Courvoisier \&
M. Mayor (Berlin: Springer), 57

\refindent
Osterbrock, D.~E. 1981, \apj, 249, 462

\refindent
Osterbrock, D.~E., \& Pogge, R.~W. 1985, \apj, 297, 166

\refindent
Peimbert, M., \& Torres-Peimbert, S. 1981, \apj, 245, 845

\refindent
Pelat, D., \& Alloin, D. 1980, \aa, 81, 172

\refindent
Peterson, B.~M. 1993, \pasp, 105, 247

\refindent
Petre, R., Mushotzky, R.~F., Serlemitsos, P.~J., Jahoda, K., \& Marshall,
F.~E. 1993, \apj, 418, 644

\refindent
Phillips, M.~M., \& Malin, D.~F. 1982, \mnras, 199, 905

\refindent
Reichert, G.~A., Branduardi-Raymont, G., Filippenko, A.~V., Mason, K.~O.,
Puchnarewicz, E.~M., \& Wu, C.-C. 1992, \apj, 387, 536

\refindent
Rose, J.~A., \& Tripicco, M.~J. 1984, \apj, 285, 55

\refindent
Sandage, A.~R., \& Tammann, G.~A. 1981, A Revised Shapley-Ames Catalog of
Bright Galaxies (Washington, DC: Carnegie Institute of Washington)

\refindent
Schmidt, M., \& Green, R.~F. 1983, \apj, 269, 352

\refindent
Shields, G.~A., \& Oke, J.~B. 1975, \apj, 197, 5

\refindent
Shields, J.~C., \& Filippenko, A.~V. 1988, \apj, 332, L55

\refindent
Shields, J. C., \& Filippenko, A. V. 1990, \aj, 100, 1034

\refindent
Shields, J. C., \& Filippenko, A. V. 1996, \aa, 311, 393

\refindent
Shuder, J.~M. 1980, \apj, 240, 32

\refindent
Shuder, J.~M., \& Osterbrock, D.~E. 1981, \apj, 250, 55

\refindent
Stauffer, J.~R. 1982, \apj, 262, 66

\refindent
St\"uwe, J.~A., Schulz, H., \& H\"uhnermann, H. 1992, \aa, 261, 382

\refindent
Tully, R.~B. 1988, Nearby Galaxies Catalog (Cambridge: Cambridge Univ. Press)

\refindent
Unger, S.~W., Pedlar, A., \& Hummel, E. 1989, \aa, 208, 14

\refindent
Veilleux, S. 1991, \apjs, 75, 357

\refindent
Veilleux, S., Cecil, G., Bland-Hawthorn, J., Tully, R.~B., Filippenko,
A.~V., \& Sargent, W.~L.~W. 1994, \apj, 433, 48

\refindent
V\'eron, P. 1979, \aa, 78, 46

\refindent
V\'eron-Cetty, M.-P., \& V\'eron, P. 1996, A Catalog of Quasars and Active
Nuclei (ESO Scientific Rep. 17)

\refindent
Walker, M.~F. 1968, \apj, 151, 71

\refindent
Weedman, D.~W. 1976, \apj, 208, 30

\refindent
Wilkes, B.~J., Schmidt, G.~D., Smith, P.~S., Mathur, S., \& McLeod, K.~K.
1996, \apj, 455, L13

\refindent
Winge, C., Peterson, B.~M., Horne, K., Pogge, R.~W., Pastoriza, M.~G., \&
Storchi-Bergmann, T. 1995, \apj, 445, 680

\refindent
Wyckoff, S., Wehinger, P.~A., Spinrad, H., \& Boksenberg, A. 1980, \apj, 240, 25

%TABLES
%\clearpage
%\begin{figure}
%%\figurenum{}
%\plotone{tables/table1_v8.ps}
%%\caption{}
%\end{figure}
%
%\clearpage
%\begin{figure}
%%\figurenum{}
%\plotone{tables/table2_v7.ps}
%%\caption{}
%\end{figure}
%
%\clearpage
%\begin{figure}
%%\figurenum{}
%\plotone{tables/table3_v3.ps}
%%\caption{}
%\end{figure}
%
%FIGURE CAPTIONS
\clearpage

\centerline{\bf{Figure Captions}}
\medskip

Fig. 1. ---
Decomposition of the \sii\ \lamb\lamb 6716, 6731 doublet.  ({\it a})
Note that in NGC 4438, the two components used to model each line have the
same intensity ratio, indicating that the electron density does not
change with velocity.  ({\it b}) Each line in NGC 3031 can be modeled with
two Gaussians, but their respective ratios for the two lines are not constant
as a consequence of variations of the electron density with velocity.  In
each case, the top panel plots the actual data in {\it solid histogram},
the individual components of the fit as light {\it solid curves}, and the final
fit as a {\it dotted curve}.  The bottom panel shows the model of each of
the lines of the doublet.

Fig. 2. ---
 ({\it a}) Decomposition of the H\al\ +
\nii\ region of NGC 4278 using the analytic narrow-line model obtained from
\sii\ \lamb 6731.  Each of the narrow lines, assumed to have identical
profiles, is modeled by two Gaussians.  We are able to extract a broad H\al\
component with FWHM $\approx$ 1950 \kms\ containing 19\% of the flux of the
total blend.  Panels ({\it b}) and ({\it c}) are the same as ({\it a}), but 
using the methods of analytic and synthetic profile scaling.
In this and in all subsequent figures, the actual data are plotted as a 
{\it solid histogram}, the individual components of the fit as light 
{\it solid curves}, and the final fit as a {\it dotted curve}.  

Fig. 3. --- 
Simulations of our automated fitting procedure to search for broad H\al\ 
(see \S\ 2.4 for details).  Broad H\al\ components with several values of 
FWHM and $f_{blend}$ were added to the high S/N, narrow-lined spectrum of 
NGC 7217 having a spectral resolution of 2.5 \AA\ at H\al\ ({\it a}).  The 
left panel illustrates the fraction of the FWHM recovered as a function of 
FWHM and $f_{blend}$.  The solid, dotted, short-dashed, long-dashed, and
dot-dashed lines correspond to $f_{blend}$ = 10\%, 20\%, 40\%, 60\%, and 80\%, 
respectively. The right panel show the fraction of $f_{blend}$ recovered as a 
function of $f_{blend}$ and FWHM.  The solid, dotted, short-dashed, and 
long-dashed lines correspond to FWHM = 20, 40, 60, and 80 \AA, respectively.  
({\it b}) Same as in ({\it a}), but for a spectral resolution of 5.0 \AA.
({\it c}) Same as in ({\it a}), but for a spectral resolution of 10.0 \AA.

Fig. 4. ---
Decomposition of the H\al\ + \nii\ region for ({\it a}) NGC 266, ({\it b}) 
NGC 315, ({\it c}) NGC 660, ({\it d}) NGC 841, ({\it e}) NGC 1052, ({\it f}) 
NGC 1068, ({\it g}) NGC 1161, and ({\it h}) NGC 1167.

Fig. 5. ---
({\it a}) Decomposition of the H\bet\ + \oiii\ \lamb\lamb 4959, 5007 region 
for NGC 1068.  The top panel shows the data in full scale, while the bottom 
panel shows an expanded view. The profile of narrow H\bet\  was constrained to 
be identical to that of the \oiii\ lines, each of which is represented by five 
Gaussians, and an additional Gaussian was introduced to model broad H\bet.  We 
fixed the FWHM of the broad component of H\bet\ to be 3210 \kms\ and its 
centroid to be redshifted by 600 \kms\ with respect to the narrow component.
({\it b}) Decomposition of the \sii\ \lamb\lamb 6716, 6731 region for NGC 
1068.  The original data and the individual Gaussian components (five for 
each line) are shown in the top panel, and the summed model profiles 
are shown in the bottom panel.

Fig. 6. ---
Decomposition of the H\al\ + \nii\ region for NGC 1358 assuming that ({\it a}) 
the profile of the \sii\ lines is identical to those of narrow H\al\ and \nii, 
({\it b}) the profile of narrow H\al\ is identical to that of \nii, but both 
different from that of \sii, and ({\it c}) the profiles of narrow H\al, \nii, 
and \sii\ are independent of each other.

Fig. 7. ---
Decomposition of the H\al\ + \nii\ region for ({\it a}) NGC 1667, ({\it b}) 
NGC 1961, ({\it c}) NGC 2273, ({\it d}) NGC 2342, ({\it e}) NGC 2639, ({\it f}) 
NGC 2655, ({\it g}) NGC 2681, and ({\it h}) NGC 2768.

Fig. 8. ---
Decomposition of the H\al\ + \nii\ region for ({\it a}) NGC 2787, ({\it b}) 
NGC 2841, ({\it c}) NGC 2911, ({\it d}) NGC 2985, ({\it e}) NGC 3031, ({\it f}) 
NGC 3079, ({\it g}) NGC 3166, and ({\it h}) NGC 3190.

Fig. 9. ---
Decomposition of the H\al\ + \nii\ region for ({\it a}) NGC 3226, ({\it b}) 
NGC 3245, ({\it c}) NGC 3367, ({\it d}) NGC 3642, ({\it e}) NGC 3718, ({\it f}) 
NGC 3884, ({\it g}) NGC 3898, and ({\it h}) NGC 3982.

Fig. 10. ---
Decomposition of the H\al\ + \nii\ region for ({\it a}) NGC 3998, ({\it b}) 
NGC 4036, ({\it c}) NGC 4051, ({\it d}) NGC 4138, ({\it e}) NGC 4143, ({\it f}) 
NGC 4168, ({\it g}) NGC 4203, and ({\it h}) NGC 4258.

Fig. 11. ---
Decomposition of the H\al\ + \nii\ region for ({\it a}) NGC 4261, ({\it b}) 
NGC 4321, ({\it c}) NGC 4388, ({\it d}) NGC 4395, ({\it e}) NGC 4438, ({\it f}) 
NGC 4450, ({\it g}) NGC 4486, and ({\it h}) NGC 4501.

Fig. 12. ---
Decomposition of the H\al\ + \nii\ region for ({\it a}) NGC 4565, ({\it b}) 
NGC 4579, ({\it c}) NGC 4594, ({\it d}) NGC 4636, ({\it e}) NGC 4651, ({\it f}) 
NGC 4698, ({\it g}) NGC 4750, and ({\it h}) NGC 4772.

Fig. 13. ---
Decomposition of the H\al\ + \nii\ region for ({\it a}) NGC 5005, ({\it b}) 
NGC 5033, ({\it c}) NGC 5077, ({\it d}) NGC 5194, ({\it e}) NGC 6500, ({\it f}) 
NGC 6951, ({\it g}) NGC 7217, and ({\it h}) NGC 7479.

Fig. 14. ---
Decomposition of the H\al\ + \nii\ region for ({\it a}) NGC 1275, ({\it b}) 
NGC 3227, ({\it c}) NGC 3516, ({\it d}) NGC 4151, ({\it e}) NGC 4235, ({\it f}) 
NGC 4639, ({\it g}) NGC 5273, and ({\it h}) NGC 5548.  Note that in order to 
display the full extent of the broader lines of the objects in this figure, 
the abscissa shows a large range of wavelengths than in the preceding figures.

%FIGURES
%\clearpage
%\begin{figure}
%\figurenum{1}
%\plotone{figs/fig1.ps}
%\caption{}
%\end{figure}
%
%\clearpage
%\begin{figure}
%\figurenum{2}
%\plotone{figs/fig2.ps}
%\caption{}
%\end{figure}
%
%\clearpage
%\begin{figure}
%\figurenum{3}
%\plotone{figs/fig3.ps}
%\caption{}
%\end{figure}
%
%\clearpage
%\begin{figure}
%\figurenum{4}
%\plotone{figs/fig4.ps}
%\caption{}
%\end{figure}
%
%\clearpage
%\begin{figure}
%\figurenum{5}
%\plotone{figs/fig5.ps}
%\caption{}
%\end{figure}
%
%\clearpage
%\begin{figure}
%\figurenum{6}
%\plotone{figs/fig6.ps}
%\caption{}
%\end{figure}
%
%\clearpage
%\begin{figure}
%\figurenum{7}
%\plotone{figs/fig7.ps}
%\caption{}
%\end{figure}
%
%\begin{figure}
%\figurenum{8}
%\plotone{figs/fig8.ps}
%\caption{}
%\end{figure}
%
%\clearpage
%\begin{figure}
%\figurenum{9}
%\plotone{figs/fig9.ps}
%\caption{}
%\end{figure}
%
%\clearpage
%\begin{figure}
%\figurenum{10}
%\plotone{figs/fig10.ps}
%\caption{}
%\end{figure}
%
%\clearpage
%\begin{figure}
%\figurenum{11}
%\plotone{figs/fig11.ps}
%\caption{}
%\end{figure}
%
%\clearpage
%\begin{figure}
%\figurenum{12}
%\plotone{figs/fig12.ps}
%\caption{}
%\end{figure}
%
%\clearpage
%\begin{figure}
%\figurenum{13}
%\plotone{figs/fig13.ps}
%\caption{}
%\end{figure}
%
%\clearpage
%\begin{figure}
%\figurenum{14}
%\plotone{figs/fig14.ps}
%\caption{}
%\end{figure}

\end{document}